\documentclass[a4paper,11pt]{article}
\usepackage{jinstpub} 
\usepackage{lineno}
\usepackage{wasysym}
\usepackage{bm}
\usepackage{subfig}
\usepackage{mathtools}

\title{\boldmath Towards a muon scattering tomography system for both low-Z and high-Z materials}

\author[a,b]{Jiahui Chen,}
\author[b,1]{Huiling Li,\note{Corresponding author.}}
\author[c,b]{Yiyue Li,}
\author[b]{Pingcheng Liu}
\affiliation[a]{University of Jinan, 336 Nanxinzhuang West Road, Jinan, China}
\affiliation[b]{Shandong Institude of Advanced Technology, 100 Panlong Road, Jinan, China}
\affiliation[c]{Shandong University, 27 Shanda Nan Lu, Jinan, China}
\emailAdd{huiling.li@iat.cn}

\abstract{Muon scattering tomography (MST) is a non-destructive technique to image various materials by utilizing cosmic ray muons as probes. A typical MST system with a two-fold track detectors is particularly effective in detecting high-$Z$ materials (e.g. nuclear materials), but difficult to recognize low-$Z$ materials (e.g. explosive materials). In this work, we present a concept of MST system to discriminate both low-$Z$ and high-$Z$ materials by extra measuring momentum of low-energy muons with a Cherenkov detector.  A toy Monte Carlo simulation to describe detector responses and multiple scatterings of a muon tracking through materials is developed for statistical tests. Based on momentum-dependent track reconstruction and image reconstruction algorithm, we evaluate separation powers of different materials in the system. The results show that momentum measurement of low-energy muons and accurate track reconstruction can improve separation power of low-$Z$ materials significantly. This may enable the MST system to detect both low-$Z$ and high-$Z$ materials with cosmic ray muons in the whole energy range.}


\begin{document}
\maketitle
\flushbottom

\section{Introduction}
Muon scattering tomography (MST) is first proposed in 2003~\cite{Borozdin2003} and explored as a non-destructive technique for imaging medium-to-large dense objects. Its principle is based on multiple Coulomb scattering (MCS) of cosmic ray muons crossing target materials. The cosmic ray muons are secondary particles produced by atmospheric interactions of primary cosmic rays, and reach the earth with an average energy of 3$\sim$4 GeV and a rate of $\sim$$\rm{1 cm^{-2}min^{-1}}$~\cite{ParticleDataGroup:2022pth}. Compared with conventional radiographic methods, such as X-rays and gamma rays, the MST technique requires no radioactive sources and can monitor well-shielded dense objects thanks to the strong penetrating power of muons. Nowadays the MST applications experience fast development and are studied in broad areas, e.g. transport control, material identification and monitoring of nuclear fuel casks or nuclear reactors~\cite{Checchia:2016vnv,Checchia2019,Bonomi:2020dmm}.

The deviation angle of MCS effect depends on atomic number $Z$ of target materials and muon momentum. In general high-$Z$ materials induce larger scattering angles than low-$Z$ materials, and low-energy muons experience larger track deflections than high-energy muons. Therefore, a typical MST system with a two-fold track detectors placed before and after the target volume is quite effective in discriminating high-$Z$ materials from medium or low-$Z$ materials, but challenging to identify low-Z materials such as explosives and illegal drugs. There are efforts to improve reconstruction performance of MST technique by appending momentum spectrometer to the typical MST design, such as with a multi-grouped momentum method by inserting additional absorption layers of known materials~\cite{Anghel:2015} or adding a fieldable Cherenkov detector with $\rm CO_{2}$ and $\rm SiO_{2}$ as radiators~\cite{Bae:2022dti,Bae:2022}. The results show that reconstruction performance benefits significantly from multi-grouped momentum information, but their discussions focus on medium-to-large dense objects. Some dedicated studies for low-Z materials are carried out by analyzing scattering and absorption information of not only cosmic ray muons but also less massive atmospheric electrons~\cite{Klimenko:2005, Blanpied:2015, Anbarjafari:2021}, which bring new insights beyond MST technique.

In this work, we discuss the effects of momentum information of low-energy muons on the MST system in discrimination of both low-$Z$ and high-$Z$ materials. We present a concept design of a MST system that consists of a solid and compact system of track detectors based on plastic scintillating fibers as well as a Cherenkov detector with a radiator of fused silica. The Cherenkov detector can classify detected muons into low-energy and high-energy events, and provides precise momentum of low-energy muons, which is different from muon spectrometers in the multi-grouped momentum method of Ref.~\cite{Anghel:2015,Bae:2022dti}. To evaluate the material separation power of the MST system, a full-chain simulation including muon transport, detector response, track reconstruction and image reconstruction is developed. Therein, the muon transport with only MCS interaction and detector response with respect to position resolution of track detectors and angle resolution of Cherenkov light are described in a toy Monte Carlo method without complex interactions and detector details involved. We implement momentum-dependent Point of Closest Approach (PoCA) to reconstruct the scattering density of materials and evaluate separation powers of low-Z and high-Z materials in the MST system. Based on the results, we suggest a strategy to discriminate both low-$Z$ and high-$Z$ materials in the MST system with cosmic ray muons in the whole energy range.

\section{Concept design of the MST system}
\begin{figure}[!h]
\centering
\hspace{1cm}
\includegraphics[scale=0.4]{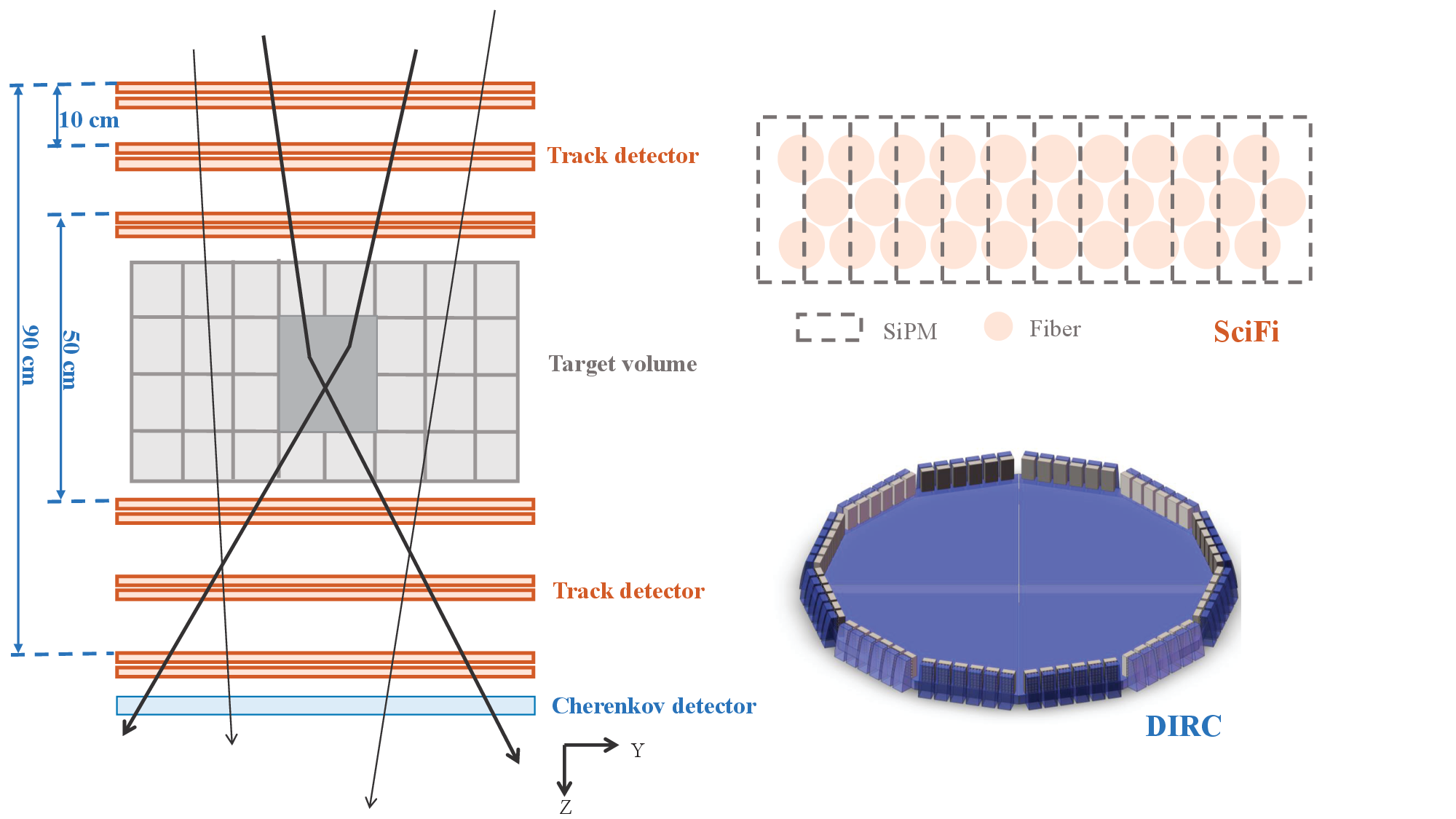}
\vspace{-0.1cm}
\caption{The concept design of the MST system consists of two track detectors and a Cherenkov detector. The area of each SciFi superlayer is set as $1\times1$ ${\rm m^{2}}$ in the following simulation and reconstruction studies. }
\label{fig:design}
\end{figure}

The concept design of a solid and compact MST system is illustrated in Fig.~\ref{fig:design}. There are two track detectors based on plastic scintillating fibers (SciFi) for measuring muon trajactories before and after the target volume to inspect. Each SciFi superlayer is set as $1\times1$ ${\rm m^{2}}$ area and placed with a separation of 10 cm. An internally reflecting imaging Cherenkov detector (DIRC) for momentum measurement of low-energy muons is located below the lower tracker. This setup geometry is applied in the following simulation and reconstruction processes for demonstration.

\subsection{Track detector}
The basic unit of SciFi detector will adopt a similar pattern as in SciFi detector for LHCb upgrade~\cite{Gruber:2020shw} that has a staggered arrangement of scintillating fibers and read out by a customized one-dimensional silicon photomultiplier (SiPM) at one end, as illustrated in Fig.~\ref{fig:design}. A detected muon position along the SiPM direction could be reconstructed by a center-of-gravity method. Its resolution is generally proportional to the fiber diameter $d$ and roughly $d/\sqrt{12}$. In the MST system, the track detector before or after the target volume should have at least three measurements of muon positions to reconstruct a reliable trajectory. Therefore, three SciFi superlayers are included in the design of a track detector and each superlayer owns two perpendicular layers of SciFi units along $X$ and $Y$ direction respectively. Meanwhile the SciFi detector can provide trigger information of cosmic ray muons.

As already demonstrated in the SciFi detector for LHCb upgrade, it is feasible to construct a $O(\rm 1m)$ long SciFi unit thanks to the long attenuation length of fibers. This could satisfy the MST requirement of large-area track detectors. Since recently a fiber of 125 $\rm \mu m$ diameter is achievable ~\cite{Kirn:2022tps}, the SciFi detector has potential to provide competitive spatial resolution even as the traditional silicon detector. A drawback brought by this high granularity of SciFi detector is a large number of readout channels, which requires a multiplexing strategy to be explored in later studies.


\subsection{Cherenkov detector}
The DIRC detector~\cite{Ratcliff:2020svc} makes use of internally reflected Cherenkov light in a solid radiator placed in the air. When the velocity $\beta$ of a cosmic ray muon exceeds the speed of light in the radiator, the Cherenkov light will be emitted in a cone with an opening angle $\theta$ around the muon trajectory that follows :
\begin{equation}
\cos \theta_{\rm ch} = \frac{1}{n\beta},\; \beta>\beta_{\rm thr}=\frac{1}{n}\;
 \label{eq:ECF}
\end{equation} 
where $n$ is the refractive index of the radiator and $\beta_{\rm thr}$ is the threshold velocity for the emission of the Cherenkov radiation. 

\begin{figure}[!t]
\begin{center}
\begin{tabular}{l}
\hspace{-1.5cm}
\includegraphics[width=0.6\textwidth]{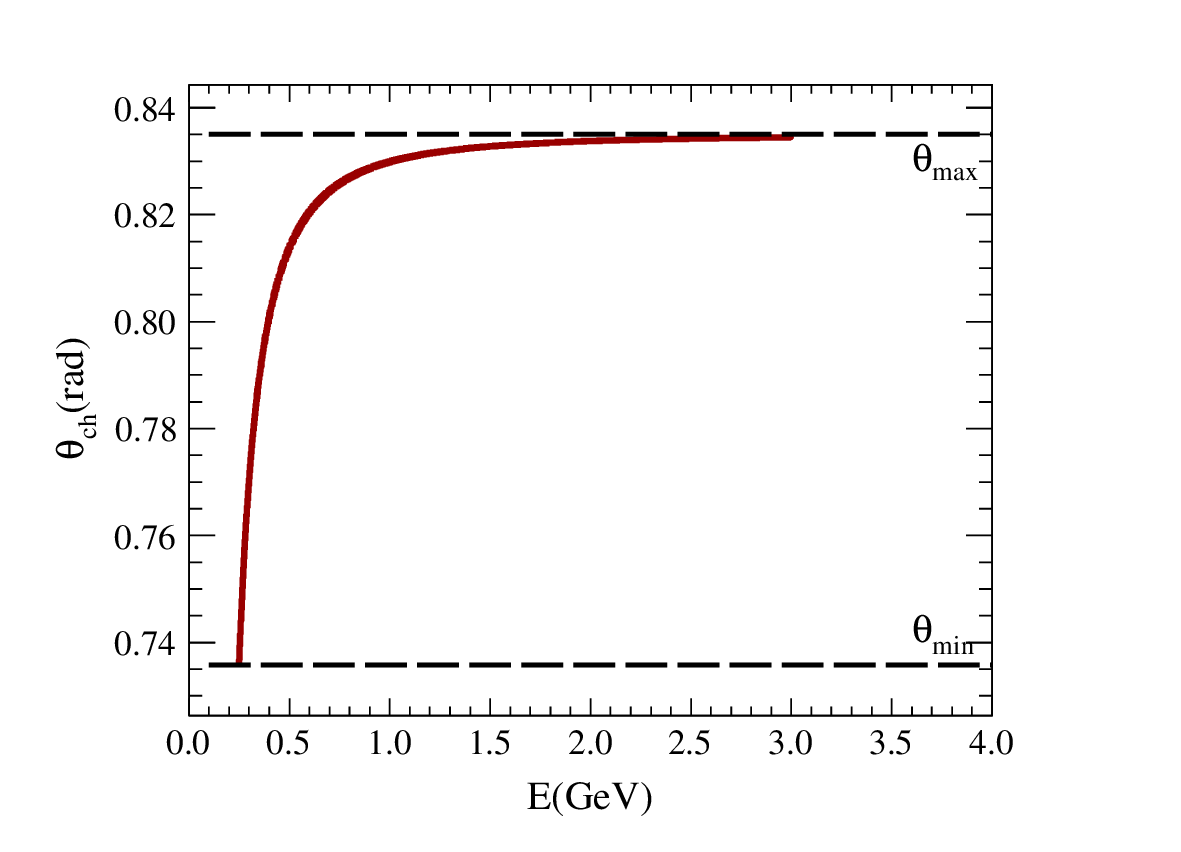}
\hspace{-1.2cm}
\includegraphics[width=0.6\textwidth]{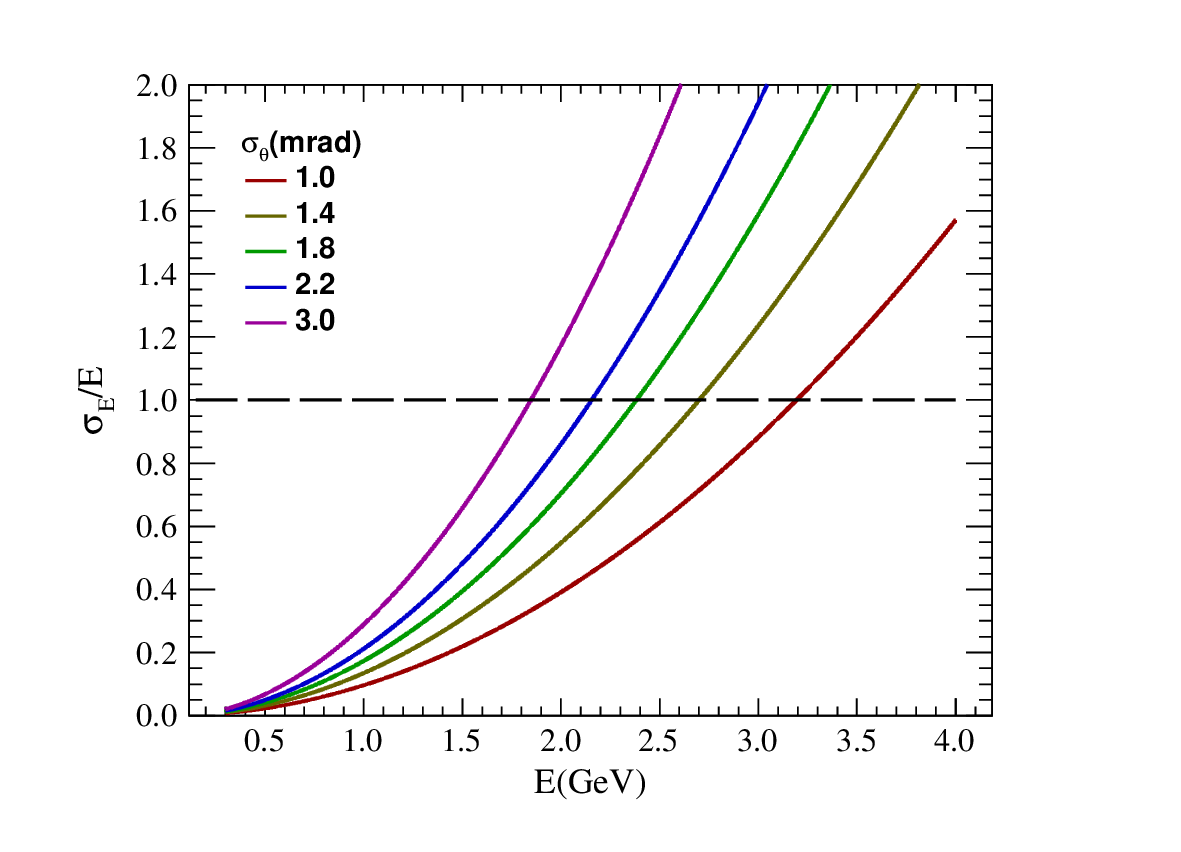}
\end{tabular}
\end{center}
\vspace{-0.6cm}
\caption{(Left) The Cherenkov angle in the fused silica for muons with different energies.  (Right) The energy resolution varies with the muon energy according to different angular resolutions.
\label{fig:resDIRC}}
\end{figure}

In this MST design, an optically transparent synthetic fused silica with a refractive index of about 1.49 is applied as the radiator. The structure of DIRC detector is shown in Fig.~\ref{fig:design} by referring to the disc DIRC of PANDA experiment~\cite{PANDA:2019unp} where the internally reflected light are detected at the edges of four quadrants by photon sensors, such as microchannel-plate photomultipliers (MCP-PMTs) or SiPMs. The Cherenkov angle in the fused silica is constrained by the muon energy as shown in the left plot of Fig.~\ref{fig:resDIRC}. It can be extracted by the hit time and position registered in photon sensors and thus the muon momentum. An advanced angle resolution between 1.2 mrad and 2 mrad can be achieved for a track in 2 cm thick radiator of PANDA disc DIRC~\cite{schonmeier2008disc,PANDA:2019unp}. Given a limited Cherenkov angle resolution of $\sigma_{\theta}$, the DIRC detector is only sensitive to muons at low energy part as demonstrated in the right plot of Fig.~\ref{fig:resDIRC}. We define an angle $\theta_{\rm lh}$ as $\theta_{\rm max}-2\sigma_{\theta}$ and then classify muon events with a reconstructed Cherenkov angle lower than $\theta_{\rm lh}$ as LE muons and those larger than $\theta_{\rm lh}$ as HE muons. A corresponding reconstruction algorithm is introduced in Sec.~\ref{sec:PoCA} for material imaging in the MST system. For the MST applications, a large-size DIRC detector is realistic since the fused silica of $O(1{\rm m})$ radius is commercially available. But the cost of the precisely made radiator and photon sensors is a drawback of the DIRC detector and may be solved as the technology develops. 

\section{Simulation of muons through MST system}\label{sec:sim}

The simulation of muon passage through the MST system mainly includes three parts, namely generator of cosmic ray muons, multiple scattering effects of muons crossing materials, and detector effects from the position resolution of the tracker and the momentum resolution of the Cherenkov detector. The energy loss of muons is neglected in this work.

\subsection{Cosmic ray muons}
The cosmic ray muon flux at sea level is generated with a modified Gaisser formula~\cite{Guan:2015vja} by taking into account the influence of muon decay and earth curvature, as shown in the following Eq.\ref{eq:generator}:
\begin{equation}
\begin{aligned}
\frac{dI}{dE d\cos{\theta}}&=0.14\left[\frac{E }{GeV}\left(1+\frac{3.64(GeV)}{E (\cos\theta^{\ast})^{1.29}}\right)\right]^{-2.7}\left[\frac{1}{1+\frac{1.1E\cos\theta^{\ast}}{115GeV}}+\frac{0.054}{1+\frac{1.1E\cos\theta^{\ast}}{850GeV}}\right]\;; \\
\cos\theta^{\ast} &=\sqrt{\frac{(\cos\theta)^2+P{_1}^2+P_2 (\cos\theta)^{P_3}+P_4 (\cos\theta)^{P_5}}{1+P{_1}^2+P{_2}+P{_4}}}\;.
\label{eq:generator}
\end{aligned}
\end{equation}
where $\theta$ is the zenith angle and $E$ is muon energy. The parameters $P_{1}$, $P_{2}$, $P_{3}$, $P_{3}$and $P_{5}$ are 0.102573, -0.068287, 0.958633, 0.0407253, 0.817285, respectively, obtained by fitting to experimental data. This modified parameterization can provide a better description of the experimental results at low energies compared with the standard Gaisser formula ~\cite{ParticleDataGroup:2022pth}.
\begin{figure}[!t]
\begin{center}
\begin{tabular}{l}
\hspace{-1.5cm}
\includegraphics[width=0.6\textwidth]{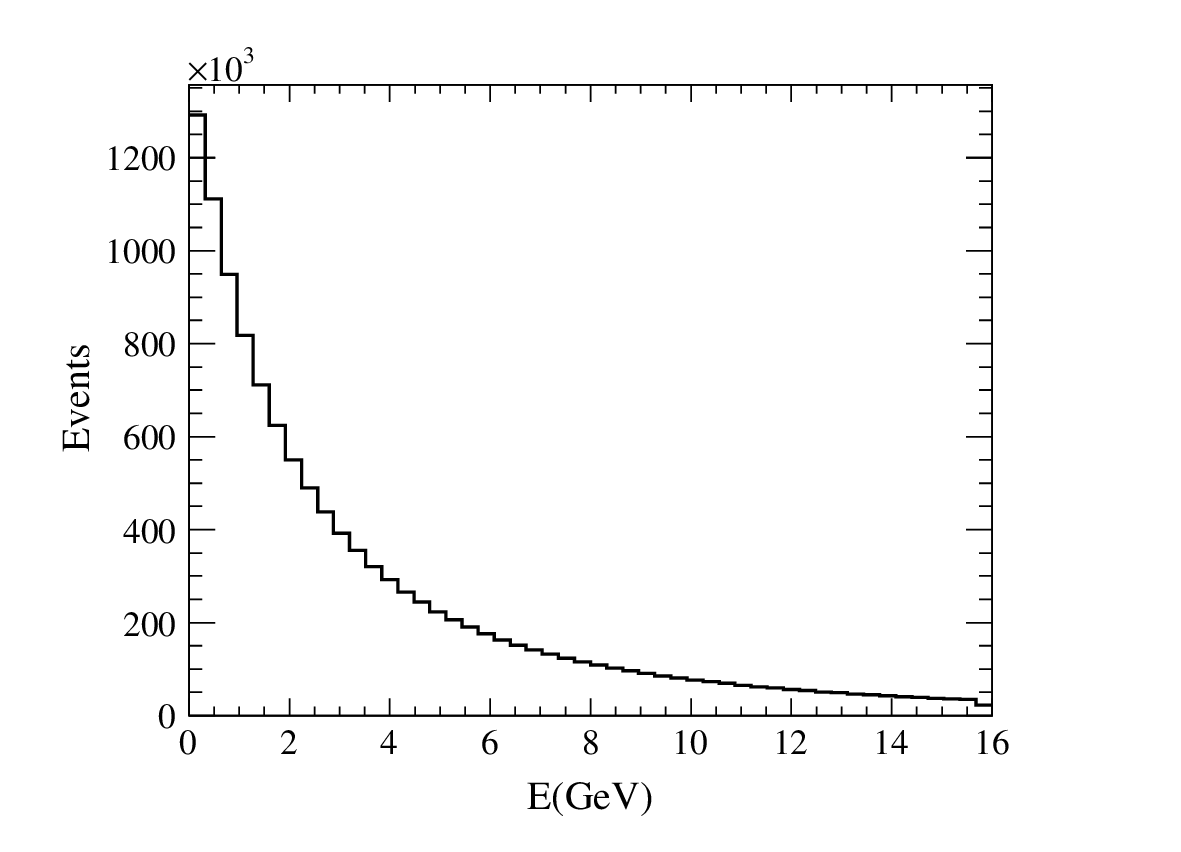}
\hspace{-1.2cm}
\includegraphics[width=0.6\textwidth]{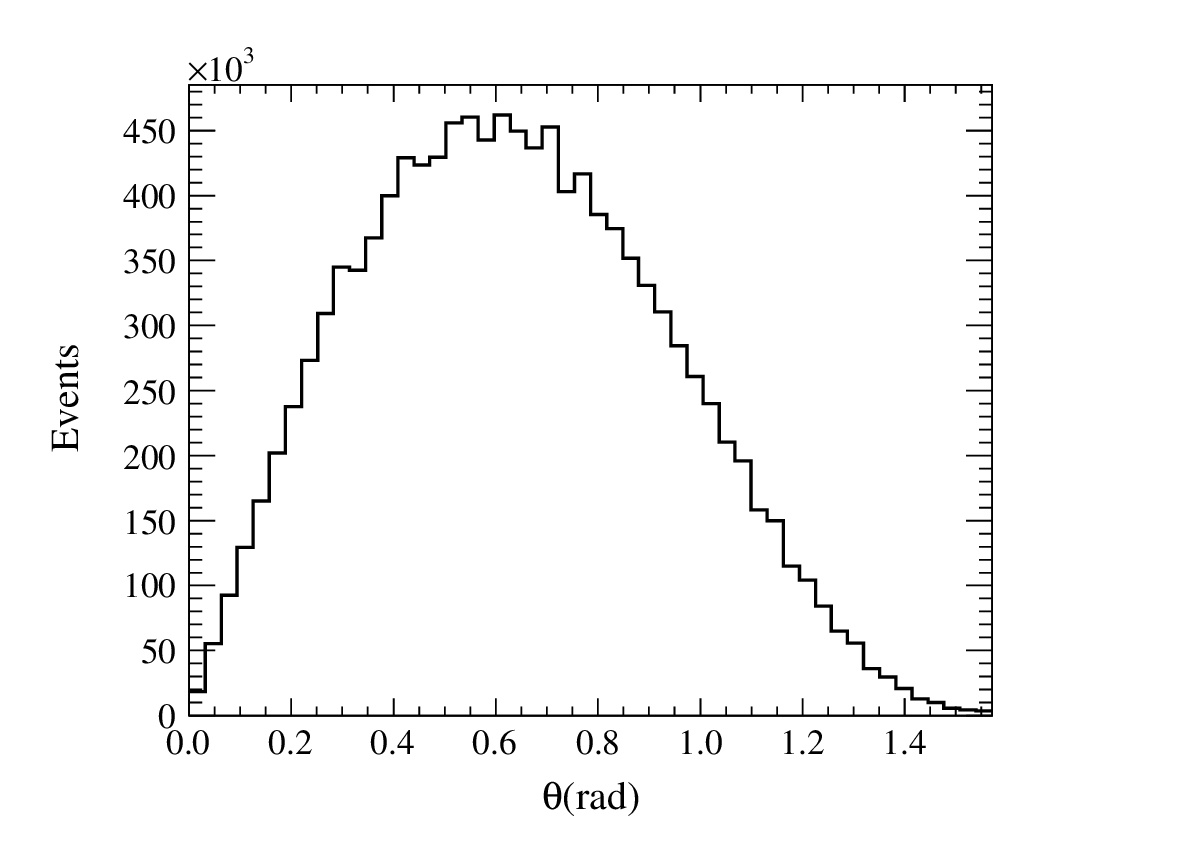}
\end{tabular}
\end{center}
\vspace{-0.6cm}
\caption{(Left) The energy spectrum and (Right) angular distribution of generated muons according to the modified Gaisser formula.}
\label{fig:muGenerator}
\end{figure}
During the simulation, a muon vertex is generated uniformly on a $3\times3{\rm~m^{2}}$ plane placed 10 m above the first tracker layer. Its energy and direction is then randomly sampled according to Eq.~\ref{eq:generator}. As shown in Fig.~\ref{fig:muGenerator} are the energy spectrum and angular distribution of generated muons, where muons dominate in the low energy range.

\subsection{Multiple Coulomb scattering effects}
Here we only simulate MCS effect of muons according to the principle of the MST system. Firstly the materials crossed by muons are sliced into thin layers to ensure small scattering angles and displacements. Then the deviation angle and displacement of MCS effect in each sliced material are described in the particle coordinate system $x^{\prime}y^{\prime}z^{\prime}$ and randomly sampled in independent projected $x^{\prime}z^{\prime}$ and $y^{\prime}z^{\prime}$ planes. Later the muon status after MCS effect is rotated into the detector coordinate system $xyz$ and the outgoing trajectory in the sliced material is estimated.

The $z^{\prime}$ axis of the particle coordinate system is defined as parallel to the muon direction, $x^{\prime}$ in the plane determined by the muon direction and $z$ axis of the detector coordinate system, and $y^{\prime}$ in the $xy$ plane of the detector coordinate system. Therefore, given the unit vector of muon incident direction in the detector coordinate system as $(u_{1},u_{2},u_{3})$, a vector $\pmb{v^{\prime}}$ in the particle coordinate system can be rotated into $\pmb{v}$ in the detector system as follows:
\begin{equation}
\begin{aligned}
U\pmb{v^{\prime}} &=\pmb{v},\;\;
\\ 
U &=
\begin{pmatrix}
\frac{u_{1}u_{3}}{\sqrt{u_{1}^{2}+u_{2}^{2}}} & -\frac{u_{2}}{\sqrt{u_{1}^{2}+u_{2}^{2}}} & u_{1} \\
\frac{u_{2}u_{3}}{\sqrt{u_{1}^{2}+u_{2}^{2}}} & \frac{u_{1}}{\sqrt{u_{1}^{2}+u_{2}^{2}}}  & u_{2} \\
\frac{u_{3}^{2}-1}{\sqrt{u_{1}^{2}+u_{2}^{2}}} & 0 & u_{3} \\
\end{pmatrix}
\;.
\end{aligned}
\label{eq:RotateMatrix}
\end{equation}

In the particle coordinate system, the MCS distribution is roughly Gaussian for small scattering angles with a projected two-dimensional form:
\begin{equation}
\begin{aligned}
  f(\theta_{\rm plane})&=\frac{1}{\sqrt{2\pi}\theta_0}exp(-\frac{\theta_{\rm plane}^2}{2\theta_0^{2}})\;; \\
  \theta_{\rm space}& \approx \sqrt{\theta_{x, \rm plane}^{2}+\theta_{y, \rm plane}^{2}}\;.
  \label{eq:MCSgaus}
  \end{aligned}
 \end{equation}
 
where $\theta_{\rm plane}$ is projected angle on $x^{\prime}z^{\prime}$ or $y^{\prime}z^{\prime}$ plane of a three-dimensional scattering angle $\theta_{\rm space}$. The $\theta_0$ is the sigma of the Gaussian distribution and estimated with:
 \begin{equation}
  \theta_0 =\frac{13.6MeV}{\beta cP}z\sqrt{\frac{L}{X_0}}[1+0.038ln(\frac{L}{X_0})]\;.\\
  \label{eq:MCStheta}
 \end{equation}
 where $X_0$ is the combined radiation length of the material, $z$ is the charge number of muons, and $L$ is the 3-dimensional pathlength along the muon incident direction through the sliced material of $\Delta z$ thickness:
 \begin{equation}
 L=\Delta z\sqrt{1+\left(\frac{u_{1}}{u_{3}}\right)^{2}+\left(\frac{u_{2}}{u_{3}}\right)^{2}}
 \end{equation} 
 
 It shows that $\theta_{0}$ of muons crossing the materials depends on the muon momentum, where low energy muons in general have larger scattering angles than high momentum muons. A toy Monte Carlo simulation following the method in Ref.~\cite{ParticleDataGroup:2022pth} is carried out to describe the deflection and displacement of muons through materials of $L$ thickness in the particle coordinate system:
\begin{equation}
\begin{aligned}
 \Delta y^{\prime}   &=c_{1} L\theta_0 /\sqrt{12}+c_{2}  L\theta_0 /2\;; \\
 \theta_{y^{\prime}}  &=c_{2} \theta_0\;.
\end{aligned}
\label{eq:MCSparticle}
\end{equation}
where $\theta_{y^{\prime}}$ and $\Delta y^{\prime}$ is the deflection angle and displacement in $y^{\prime}z^{\prime}$ plane and so is the projection in $x^{\prime}z^{\prime}$ plane. $c_1$ and $c_2$ are independent random variables following the normal distribution.

\begin{figure}[htbp]
\centering 
\hspace{-2cm}
\includegraphics[scale=0.8]{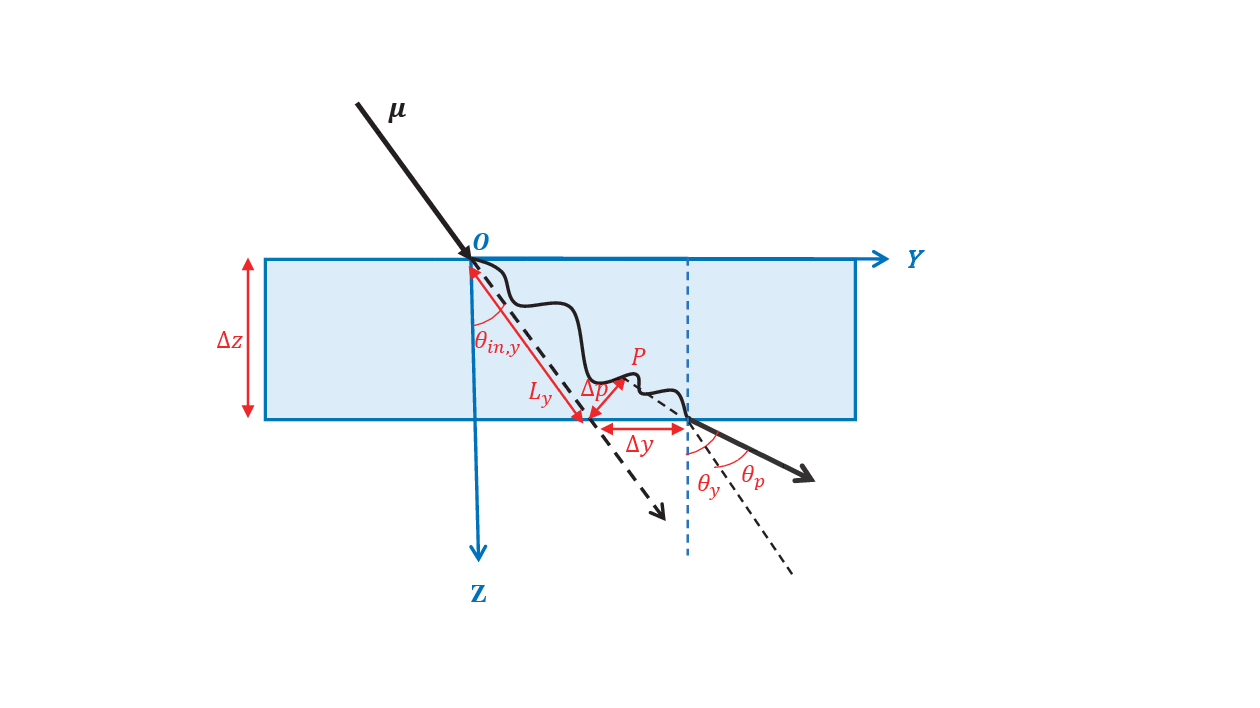}
\vspace{-2cm}
\caption{Deflection and displacement in sliced material of a muon trajectory in two-dimensional detector coordinate system. The displacement $\Delta p$ and the deviation angle $\theta_{p}$ are induced by the MCS effect and rotated from the particle coordinate system.}
\label{fig:MCSsim} 
\end{figure}

\begin{figure}[htbp]
\centering 
\includegraphics[scale=0.5]{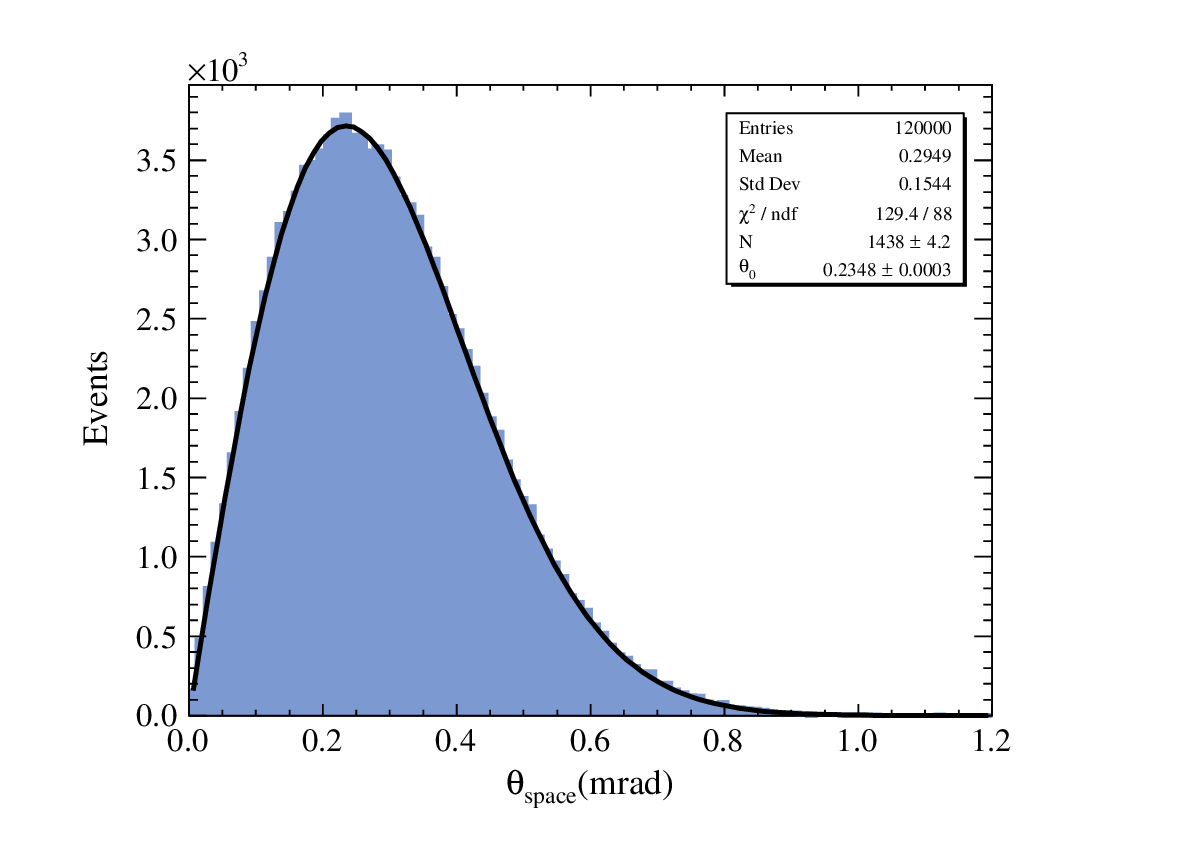}
\caption{The deviation angle distribution of the 3 GeV muon trajectories crossing through the SciFi layer vertically. Here the thickness of the layer is taken as $1420 \rm \mu m$ and the radiation length is $33.2 \rm cm$. }
\label{fig:MSdeteAng} 
\end{figure}

Then the muon position and direction after the MCS effect in the particle coordinate system are rotated into the detector coordinate system following Eq.~\ref{eq:RotateMatrix}. As illustrated in Fig.~\ref{fig:MCSsim}, this position projected in the $yz$ plane is denoted as $P$. $\Delta p$ and $\theta_{p}$ are resulted from the MCS effect, which are different from the $\Delta y^{\prime}$ and $\theta_{y^{\prime}}$ after coordinate transformation. We assumed that the muon direction at $P$ is the same as the direction of muon exit position and thus the deviation angle at the exit point is approximately the same as $\theta_{p}$. Then the muon outgoing trajectory on the $yz$ plane can be estimated with the following:
\begin{equation}
\begin{aligned}
 \Delta y &= \frac{\Delta p \cos\theta_{p}}{\cos\theta_{y}},\;\; \theta_{y} = \theta_{in,y}+\theta_{p}\;.
 \label{eq:MCSdetector}
\end{aligned}
\end{equation}
where $\theta_{in,y}$ is the angle of incident trajectory. This formula is only valid for thin sliced materials. Meanwhile, the outgoing trajectory on the $xz$ plane can be processed in a similar way as that on the $yz$ plane. We validate the toy Monte Carlo process of MCS effect by fitting the $\theta_{\rm{space}}$ of sampled muon events with Eq.~\ref{eq:MCSgaus} as shown in Fig~\ref{fig:MSdeteAng}, where the fitted value of $\theta_{0}$ is consistent with the theoretical value estimated from Eq. 3.4.

%

\subsection{Detector Effects}
In practice, the reconstructed data of the MST system used in image reconstruction are convolution of true track information of muons and detector effects. Therefore we further add the spatial resolution of track detectors on muon trajectories after MCS effect and angular resolution of DIRC detector on true muon energy. 

\begin{itemize}
\item For each SciFi layer of track detectors, a reconstructed position is obtained from the exit position after MCS effect smeared by a Gaussian distribution with exit position as mean and spatial resolution as sigma. We consider three cases of spatial resolution according to recent researches of SciFi detector, i.e. in cosmic ray tracker~\cite{Zhang:2022ehk}, LHCb experiment~\cite{Gruber:2020shw} and AMS-100 experiment~\cite{Kirn:2022tps} as shown in Tab.~\ref{tab:scifiRes}. Here the spatial resolution is estimated as $d/\sqrt{12}$ with $d$ denoting the fiber diameter.

\begin{table}[htbp]
 \centering
  \caption{Parameters of three SciFi detectors proposed in recent research.}
 \smallskip
 \begin{tabular}{lccc}
  \hline 
application &  fiber diameter $(\mu m)$   & spatial resolution $(\mu m)$  & layer thickness  $(\mu m)$ \\ 
    \hline
cosmic ray tracker   &  1000   &  300   & 2906 \\ 
    LHCb         &  250   &  72    & 1305  \\ 
    AMS-100      &  125   &  36    & 1420  \\ 
    \hline
  \end{tabular}
  \label{tab:scifiRes}
\end{table}	

\item The Cherenkov detector is used to detect the muon momentum after the SciFi detector in MST system. Since the induced Cherenkov angle $\theta_{\rm ch}$ depends on the velocity $\beta$ of the incident muon, we can estimate its corresponding energy resolution $\sigma_{E}$ after error propagation of angular resolution $\sigma_{\theta}$ as shown in the following:
\begin{equation}
 \sigma_{E}  =\frac{E\beta}{1-\beta^2} \sigma_{\beta},\;\; \sigma_{\beta} =\beta\sigma_{\theta}\tan\theta_{\rm ch}\;.
\end{equation}
Then reconstructed energy by DIRC detector for each muon is randomly sampled from a Gaussian distribution with $E$ as mean and $\sigma_{E}$ as sigma,which is used to classify LE and HE muons in the following reconstruction according to the theoretical energy value of Cherenkov angle $\theta_{\rm lh}$ .
\end{itemize}

\section{Reconstruction methods}
The MST technique aims to reconstruct the spatial distribution of the linear scattering density $\lambda$ within the target volume, where $\lambda$ is defined as the reciprocal of radiation length $1/X_{0}$ according to Eq.~\ref{eq:MCStheta} after neglecting the logarithmic term:
\begin{equation}
	\lambda=\frac{1}{X_0}\approx \left( \frac{\beta cP}{13.6MeV}\right)^{2} \frac{\theta_{0}^{2}}{Lz^{2}}.
	\label{eq:LSD}
\end{equation}
Typically such reconstruction is realized with two steps of muon track reconstruction and image reconstruction of target materials by utilizing simulated events from Sec.~\ref{sec:sim}. Since momentum of low-energy muons can be measured by DIRC detector, we introduce Kalman filter algorithm for track reconstruction and momentum-dependent Point of Closest Approach (PoCA) for image reconstruction in the following. 

\subsection{Kalman filter algorithm}
The Kalman filter (KF) algorithm is widely used in high energy physics experiments for track reconstruction. It determines an optimal recursive estimator of particle track state dynamically from one layer to the next. Compared with traditional global least-squares fit, it can naturally take multiple scattering and energy loss of particles through materials into account. In this work, only MCS effect of materials is included, and the muon tracks before and after the target volume are reconstructed separately.

Based on the equations and notations of Ref.~\cite{Fruhwirth:1987fm}, the KF algorithm generally operates in two procedures of filtering and smoothing to reconstruct a muon track. The filtering starts from the first hit of a muon track and predicts the status vector of the next hit with a straight-line model following:
\begin{equation}
\begin{aligned}
     \bm{x_{k}} &=F_{k-1} \bm{x_{k-1}}+\bm{\omega_{k-1}}\;,
     \\
     \bm{x_{k-1}} &=
\begin{pmatrix}
u_{k-1} \\ t_{k-1}
\end{pmatrix}
,\;\;
F_{k-1} =
\begin{pmatrix}
1 & d \\
0 & 1
\end{pmatrix}
\;.
\end{aligned}
\end{equation}
Here a muon track is reconstructed with two independent projected tracks on $xz$ and $yz$ planes. For instance, on $xz$ projection, the state vector $\bm{x_{k-1}}$ and propagation matrix $F_{k-1}$ are defined by a two-dimensional straight line model. $u_{k-1}$ and $t_{k-1}$ of the state vector is x-position and slope of the track on layer $k-1$. $d$ is the distance of two tracker planes. The $\bm{\omega_{k-1}}$ is process noise from MCS effect, which contributes to the covariance matrix $C_{k}^{k-1}$ of the predicted hit on the next tracker layer:

\begin{equation}
\begin{aligned}
C_{k}^{k-1} &= F_{k-1}C_{k-1}F_{k-1}^{T}+Q_{k-1}\;.
\\
Q_{k-1} &= 
\begin{pmatrix}
\theta_{0}^{2}t^{2} \Delta z^{2} /3 & \theta_{0}^{2} t^{2} \Delta z/2\\
 \theta_{0}^{2}t^{2}\Delta z /2  &  \theta_{0}^{2}t^{2}
\end{pmatrix}
\end{aligned}
\end{equation}

where $\theta_0$ is the root mean square of the multiple scattering angles as defined in Eq.~\ref{eq:MCStheta}, $\Delta z$ is the material thickness along $z$ axis of SciFi detector and $t=\sqrt{1+t_{x}^2+t_{y}^2}\sqrt{1+t_{x}^2}$ in $xz$ plane. $Q_{k-1}$ is a symmetric matrix derived from $\omega_{k-1}$~\cite{Wolin:1992ti, Rainer:1998}.

The predicted state vector is later filtered and updated in light of the measured position vector $\bm{m_{k}}$.The measurements are linear functions of track state: 
\begin{equation}
\begin{aligned}
\bm{m_{k}} &= H_{k}\bm{x_{k}}+\epsilon_{k}\;,
\\
\bm{m_{k}} &=
\begin{pmatrix}
r_{k}^{x} \\ 0
\end{pmatrix}
,\;\;
H_{k} = 
\begin{pmatrix}
1 &  0 \\
0 & 0 
\end{pmatrix}
\;.
\end{aligned}
\end{equation}
where $r_{k}^{x}$ is the measured position along $x$ axis on $k$ layer. $\epsilon_{k}$ is from the measurement uncertainty of the track detector and determines $V_{k}$ matrix in covariance matrix of filtered state vector.
\begin{equation}
V_{k} =
\begin{pmatrix}
\sigma_{x}^{2} && 0 \\
0 && 0
\end{pmatrix}\;.
\end{equation}
where $\sigma_{x}$ is spatial resolution of SciFi detector. In the forward filtering procedure, the estimated state on a given layer of the track detector takes no measured information on farther layers. Therefore when the filtering is finished, a smoothing procedure steps back up the track from the bottom and refines the estimation of track state on each layer. Here the smoothing method also follows that in Ref.~\cite{Fruhwirth:1987fm}. 

The seeding state $\bm{x_{0}}$ of the filtering is determined by measured values on the first SciFi tracker layer, while the seeding of smoothing applies the filtered track state on the last layer. Regarding the seed of covariance matrix $C_{0}$ its elements are set as $10^{6}$ at the beginning of filtering and enlarged before each filtering and smoothing processes in later iterations, which will stop until the change of chi-square values of adjacent processes is small. 


\begin{figure}[t!]
\begin{center}
\begin{tabular}{c}
\includegraphics[width=0.5\textwidth]{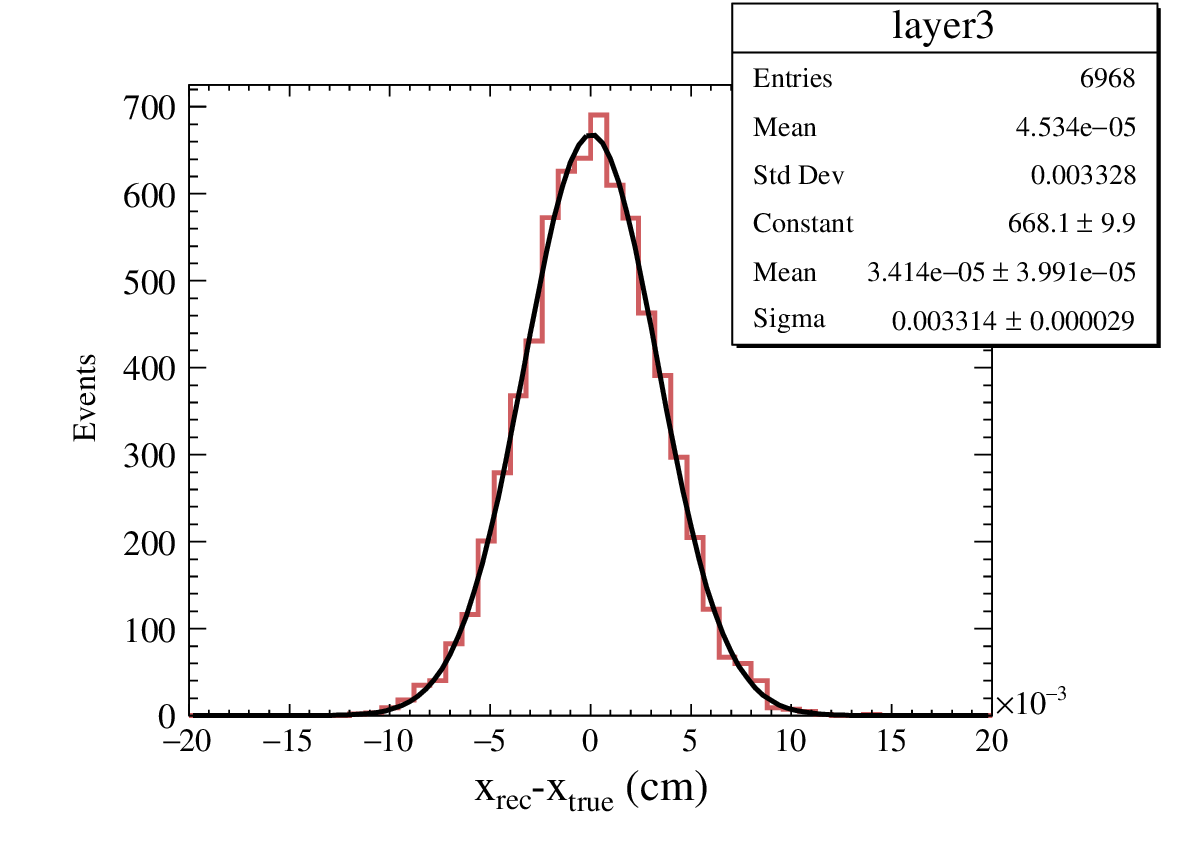}
\hspace{-0.2cm}
\includegraphics[width=0.5\textwidth]{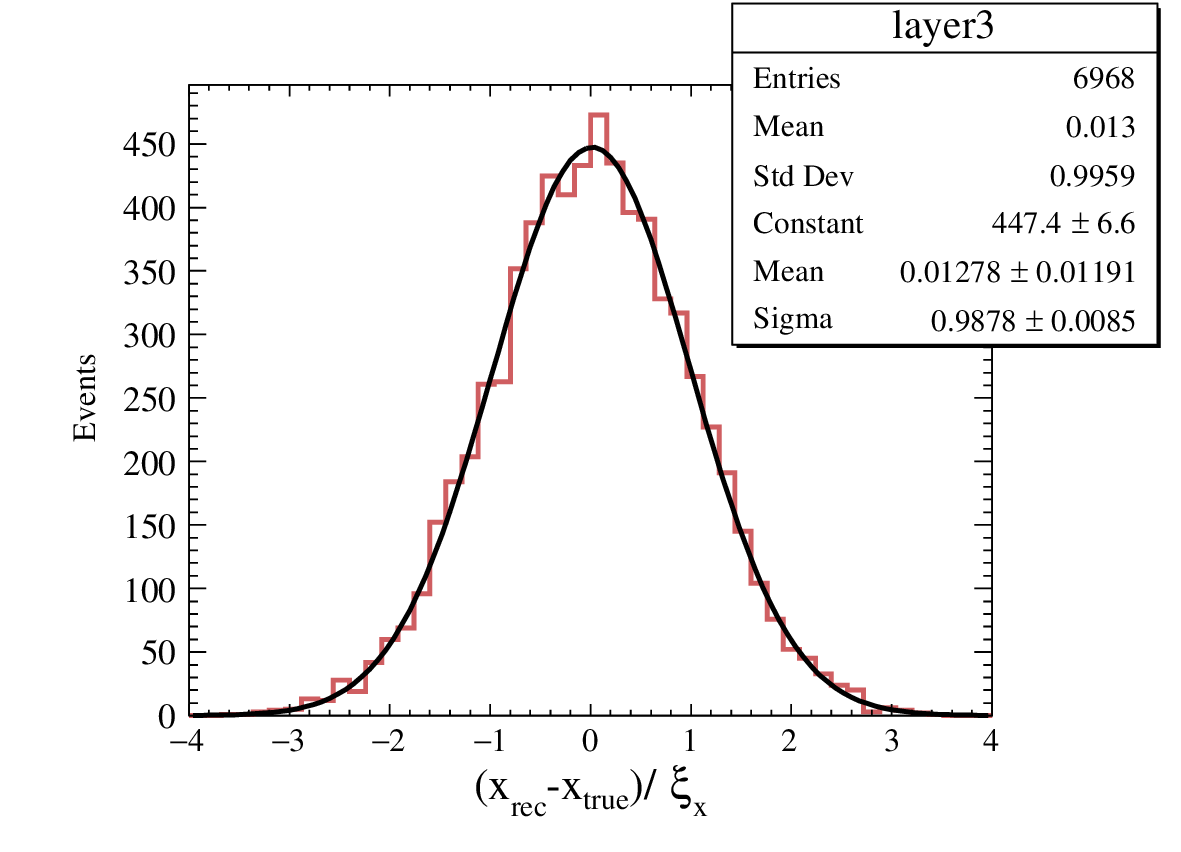}
\\
\includegraphics[width=0.5\textwidth]{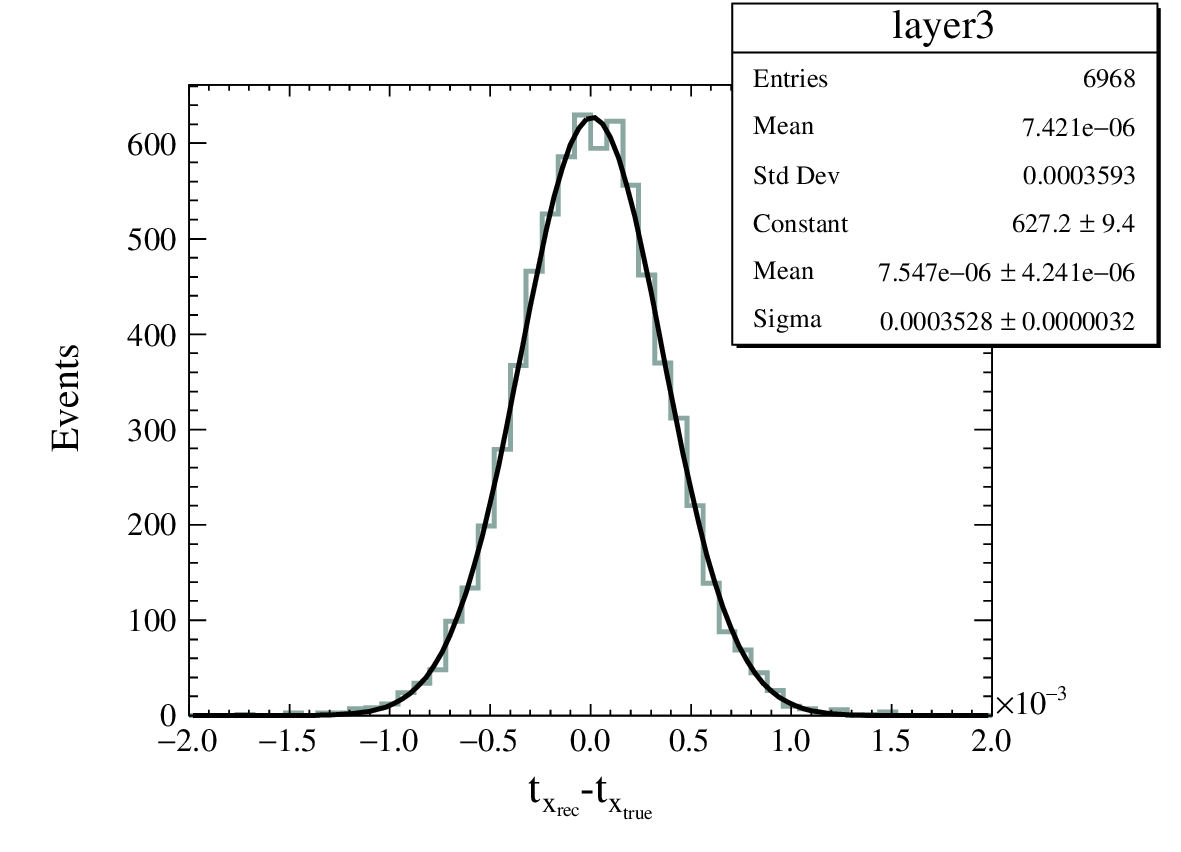}
\hspace{-0.2cm}
\includegraphics[width=0.5\textwidth]{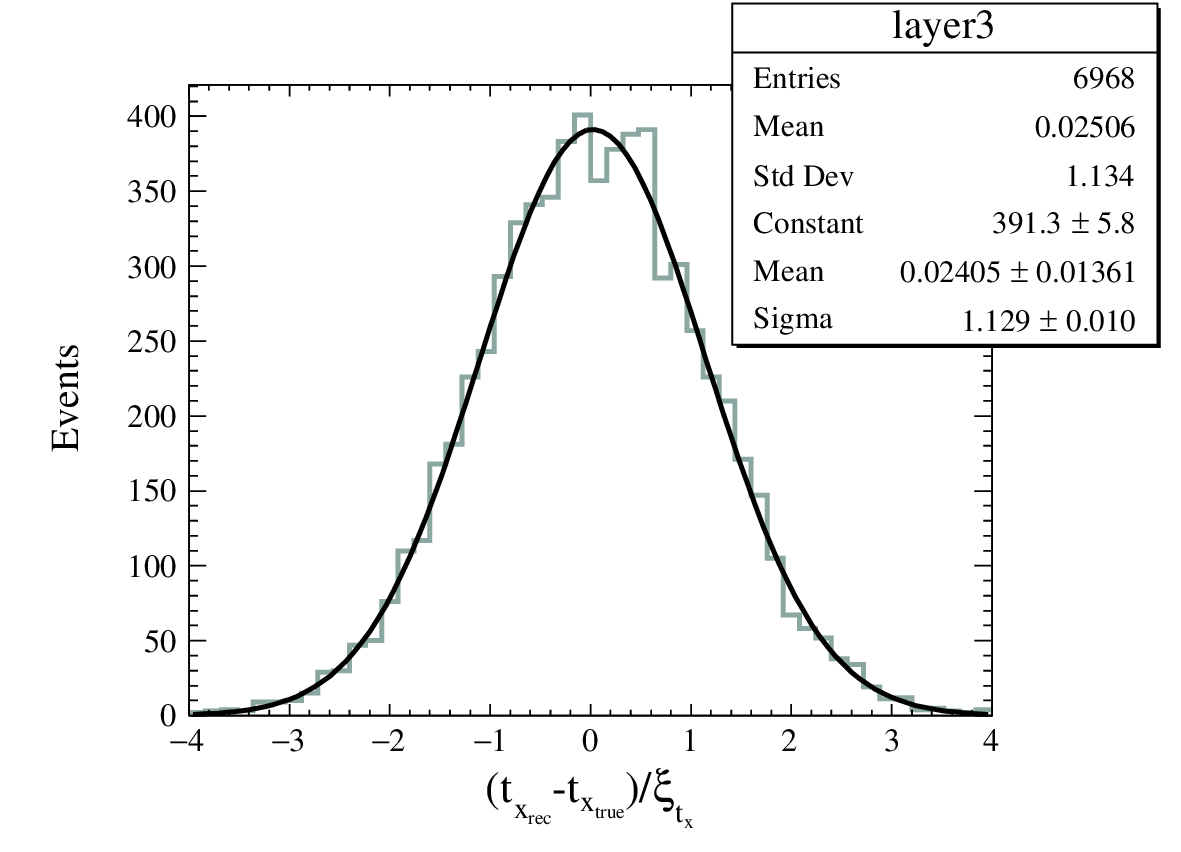}
\end{tabular}
\end{center}
\vspace{-0.5cm}
\caption{The residual (left) and pull distribution (right) of reconstructed track parameters on third layer of the upper tracker, which are fitted well with Gaussian distributions. Here $\xi_{\rm x}$ and $\xi_{t_{\rm x}}$ are the variances of $x_{\rm rec}$ and $t_{x_{\rm rec}}$ obtained from the KF track fitting. }
\label{fig:KFresandpulldis}
\end{figure}

\begin{figure}[htbp]
\centering 
\includegraphics[scale=0.45]{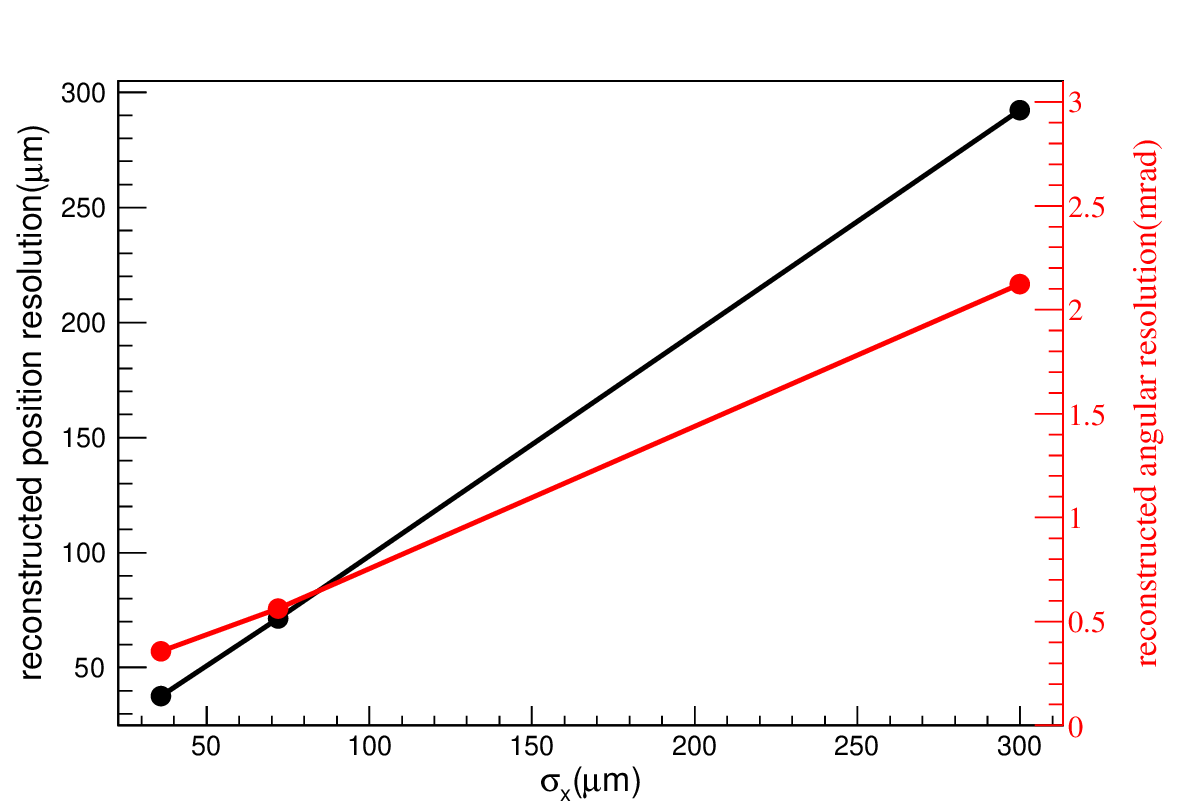}
\caption{Reconstucted position resolutions and angular resolutions at the upper edge of the target volume by using KF algorithm in cases of $36 \rm{\mu m}$, $72 \rm{\mu m}$ and $300 \rm{\mu m}$ resolutions of track detector.}
\label{fig:Kfreconstruction} 
\end{figure}

The residual and pull distributions of reconstructed parameters are checked for each layer. As an illustration, Fig.~\ref{fig:KFresandpulldis} shows the reconstructed $x_{\rm rec} $ and $t_{x_{\rm rec}}$ results at the third layer of upper track detector for a case of $36{\rm \mu m}$ position resolution. We also evaluate the reconstructed position resolutions and angular resolutions at the upper edge of the target volume with different spatial resolutions of SciFi detector as shown in Fig.~\ref{fig:Kfreconstruction}. The $O(10\rm{\mu m})$ resolution of track detectors enables an angular resolution of $O(0.1 \rm{mrad})$ in the MST geometry. Additionally, we point out that the KF algorithm is quite useful to determine deviation angles in tracker materials, such as in PTF method to separate atmospheric muons and electron~\cite{Anbarjafari:2021} and multi-grouped momentum method with known material layers~\cite{Anghel:2015}.

\subsection{Momentum-dependent PoCA}\label{sec:PoCA}
A muon traversing target volume might experience multiple scatterings on materials. This picture is simplified by the typical PoCA method which assumes that a muon is scattered at a single point. This scattering point can be determined as the midpoint of the shortest line segment simultaneously perpendicular to incoming and outgoing muon tracks in three-dimensional space~\cite{Zeng:2019ewo}. The target volume is divided into many voxels. A scattering angle $s_{i}$ is then assigned to the single voxel located by the scattering point of a muon event:
\begin{equation}
s_{i}^{2} = \frac{1}{2}\left[\left(\theta_{x_{i}}^{\rm out}-\theta_{x_{i}}^{\rm in}\right)^{2}+\left(\theta_{y_{i}}^{\rm out}-\theta_{y_{i}}^{\rm in}\right)^{2}\right]\;.
\end{equation}
where the material thickness of MCS effect is taken as the voxel size $L_{k}$. Given multiple muon events with scattering points located in the same voxel, the linear scattering density of this voxel can be collectively estimated by their scattering angles. 

According to Eq.~\ref{eq:LSD}, for the case without momentum measurement, a typical treatment is to use an average momentum for all muon events in the estimation as the following:
\begin{equation}
\lambda^{\rm ave}_{k}=\left( \frac{\bar{\beta} c\bar{P}}{13.6MeV}\right)^{2}\frac{1}{N_{k}L_{k}} \sum_{i=1}^{N_{k}} {s_{i}^{2}}\;.
\label{eq:pocaAveE}
\end{equation}
where $\bar{P}$ is the average energy of about 3.5 GeV according to the muon generator and $\frac{1}{N_{k}} \sum_{i=1}^{N_{k}} {s_{i}^{2}}$ is an approximation of $\theta_{0}^{2}$ since the mean value of scattering angle distribution in Eq.~\ref{eq:MCSgaus} is zero.  The fixed $\bar{P}$ for each muon event overweights the scattering contribution of low energy muons, which results in the reconstructed linear scattering density biased from the theoretical value of the material. In the extreme case of precise momentum measured for all muons, the reconstructed result can be estimated with:
\begin{equation}
\lambda_{k}^{\rm true} = \frac{1}{N_{k}L_{k}} \sum_{i=1}^{N_{k}} {\left( \frac{\beta_{i} c P_{i}}{13.6MeV}\right)^{2}s_{i}^{2}}\;.
\label{eq:pocaTrueE}
\end{equation}

\begin{figure}[htbp]
\centering 
\includegraphics[scale=0.5]{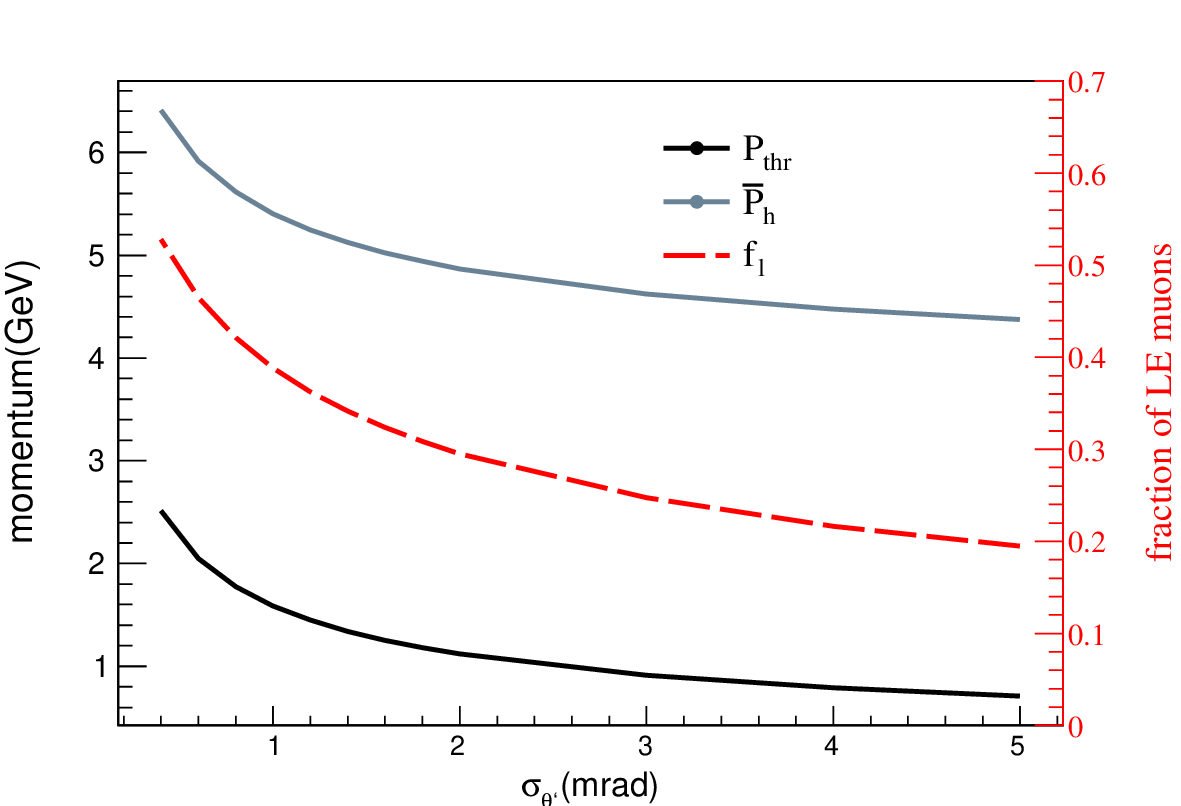}
\caption{The $P_{\rm thr}$, $\bar{P}_{\rm h}$ and the fraction of LE muons $f_{l}$ used in the segmented momentum-dependent PoCA method varies with the angular resolution $\sigma_{\theta}$ of the Cherenkov detector.}
\label{fig:momedependPoCA} 
\end{figure}

In the proposed MST system, the momentum of low-energy muons can be measured by the DIRC detector, while a high-energy event cannot be reconstructed well as demonstrated in the right plot of Fig.~\ref{fig:resDIRC}. Therefore we apply a segmented momentum-dependent PoCA method that muon events in a voxel are classified into two categories of LE and HE muons as the following:
\begin{equation}
\begin{aligned}
\lambda_{k}^{l} &= \frac{1}{N_{k}^{l}L_{k}} \sum_{i=1}^{N_{k}^{l}} {\left( \frac{\beta_{i} c P_{i}}{13.6MeV}\right)^{2}s_{i}^{2}}\;\;\;&P_{i}\leq P_{\rm thr}\;,\\
\lambda_{k}^{h}&=\left( \frac{\bar{\beta}_{h} c\bar{P}_{h}}{13.6MeV}\right)^{2}\frac{1}{N_{k}^{h}L_{k}} \sum_{i=1}^{N_{k}^{h}} {s_{i}^{2}}\;\;\;&P_{i}> P_{\rm thr}\;.
\end{aligned}
\label{eq:LSDsegMom}
\end{equation}
Here $N_{k}^{l}$ and $N_{k}^{h}$ is the number of LE and HE muon events in the $k^{\rm th}$ voxel, respectively. The value of $P_{\rm thr}$ depends on the angular resolution of the DIRC detector and is determined as the muon momentum at the Cherenkov angle $\theta_{\rm lh}$. $\bar{P}_{\rm h}$ is defined as the average momentum of HE muons, which is used to estimate linear scattering density collectively. As shown in Fig.~\ref{fig:momedependPoCA}, the values of $P_{\rm thr}$, $\bar{P}_{\rm h}$ and the fraction of LE muons $f_{\rm l}$ are obtained from the muon generator for different angle resolutions. $P_{\rm thr}$ and $\bar{P}_{\rm h}$ decrease when the angular resolution gets worse. The LE muons account for quite a proportion of cosmic ray muons, which is about $30\%$ for an angular resolution of 2 mrad.

\section{Material discrimination}
\begin{figure}[htbp]
\centering 
\includegraphics[scale=0.5]{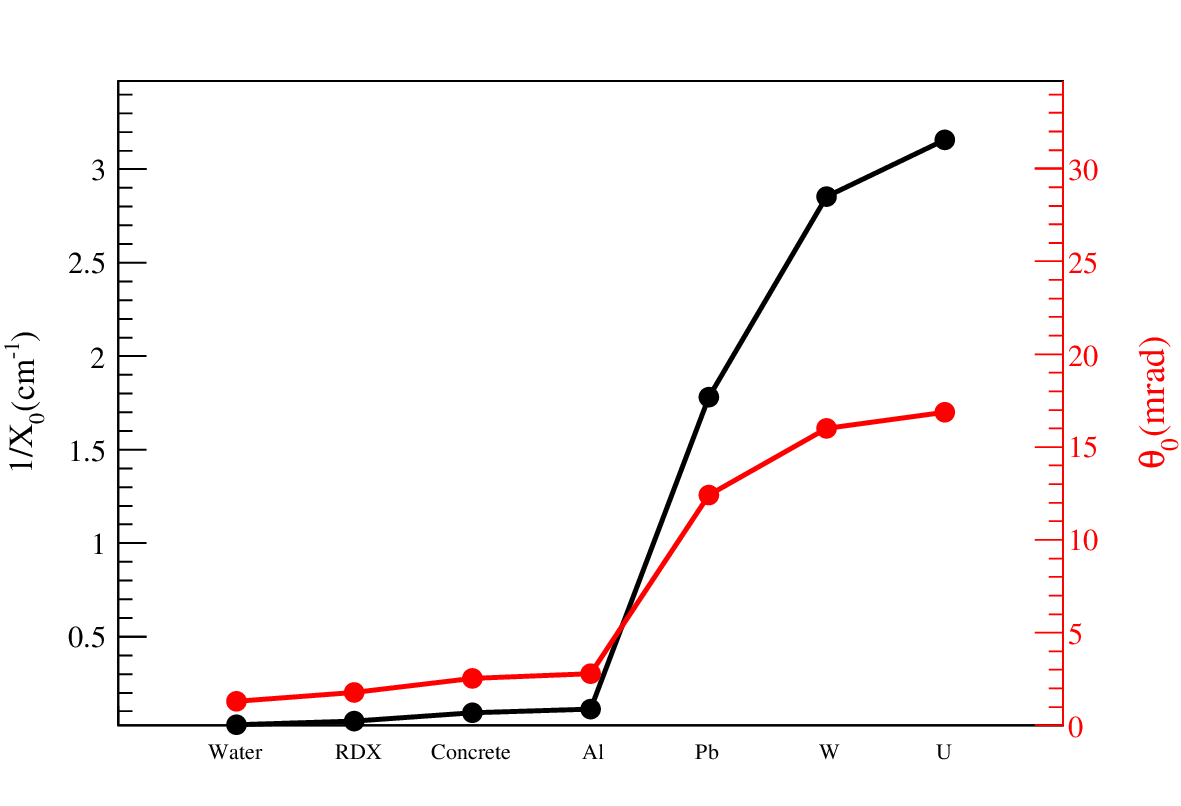}
\caption{\label{fig:LSDmaterial} Theoretical linear scattering density $1/X_0$ of different materials (black points) and $\theta_0$ of MCS effect of 3.5 GeV muons crossing through different materials with a thickness of 5 cm (red points). }
\end{figure}

The materials of $Z$ values ranging from low to high are illustrated in Fig.~\ref{fig:LSDmaterial}, where the linear scattering density of a low-$Z$ material can be about two orders of magnitude lower than that of a high-$Z$ material. In statistical tests, a cube of a single isolated material with a dimension of 10 cm  is placed in the center of the target volume. We select muon events with reconstructed tracks within the MST system geometry and further remove non-scattered events with deviation angles of incoming and outgoing tracks smaller than a selected threshold that depends on the angular resolution of reconstructed tracks. Then the PoCA reconstruction is performed with simulated data in the target volume divided into many $5\times5\times5$ ${\rm cm^{3}}$ voxels.

\begin{figure}[t!]
\begin{center}
\begin{tabular}{c}
\includegraphics[width=0.55\textwidth]{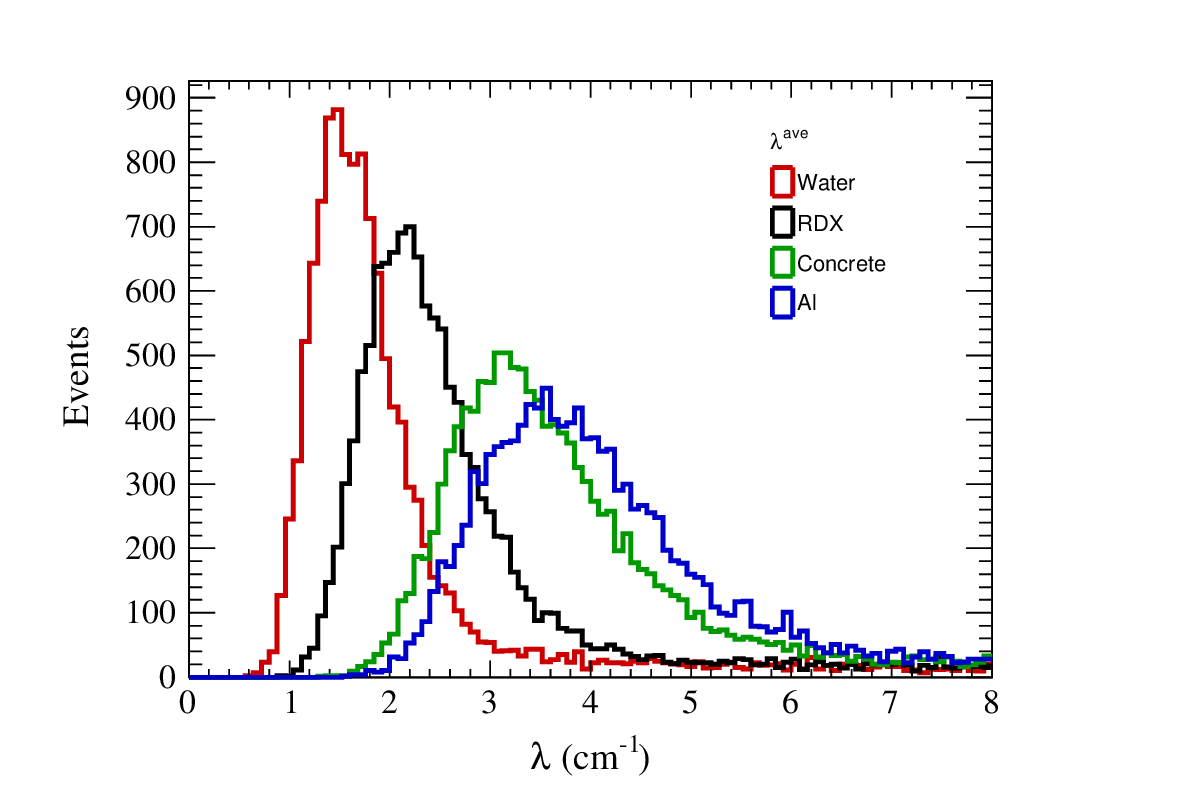}
\hspace{-1.2cm}
\includegraphics[width=0.55\textwidth]{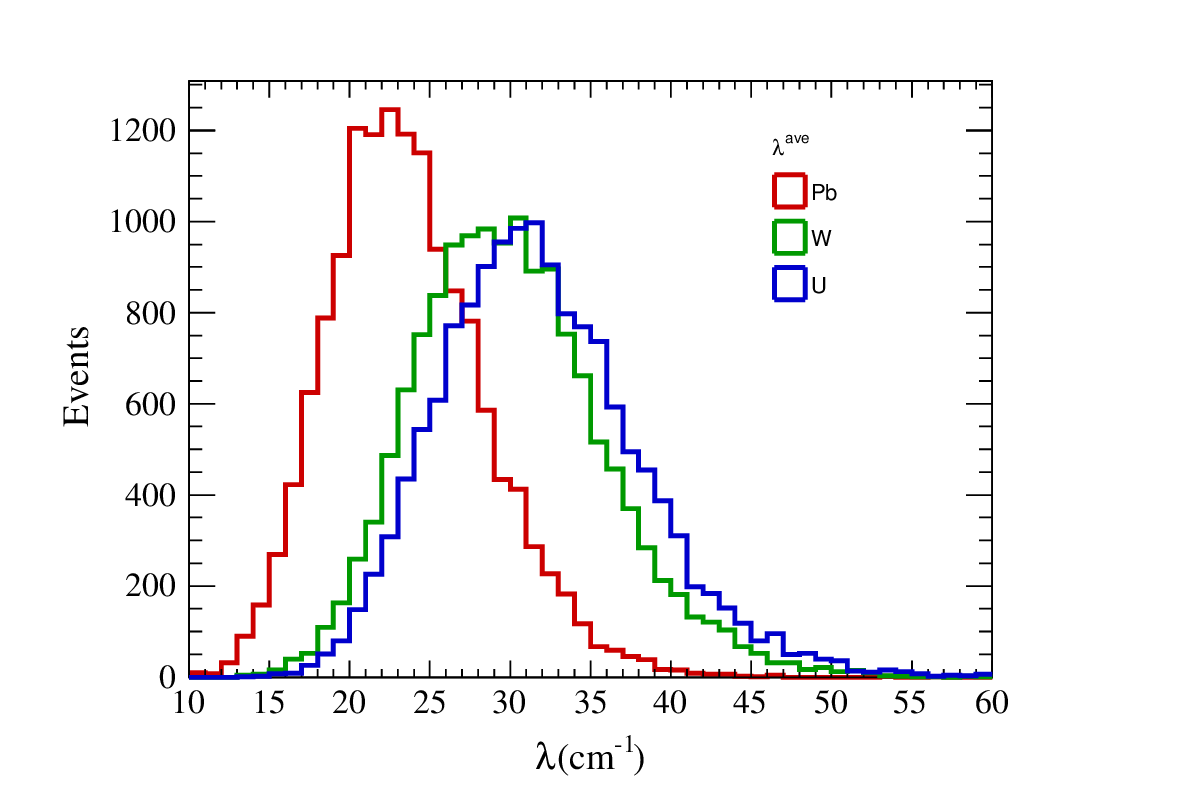}
\\
\includegraphics[width=0.55\textwidth]{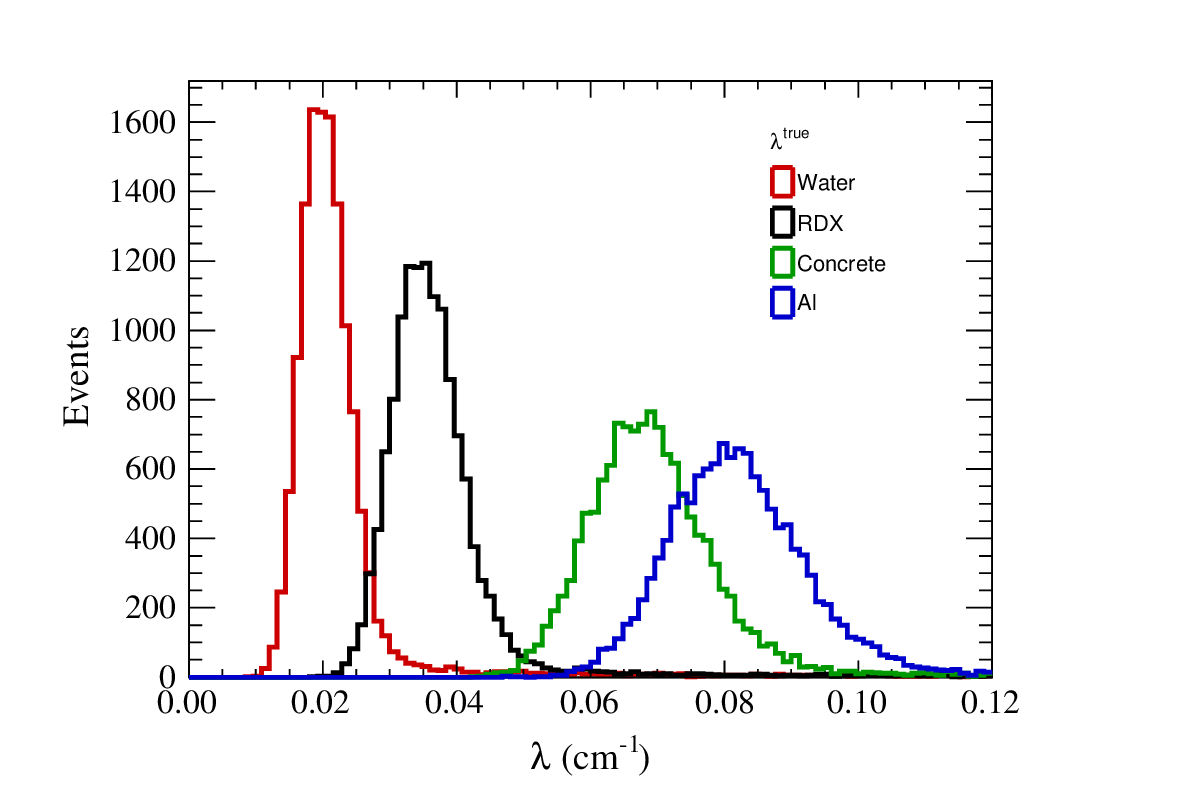}
\hspace{-1.2cm}
\includegraphics[width=0.55\textwidth]{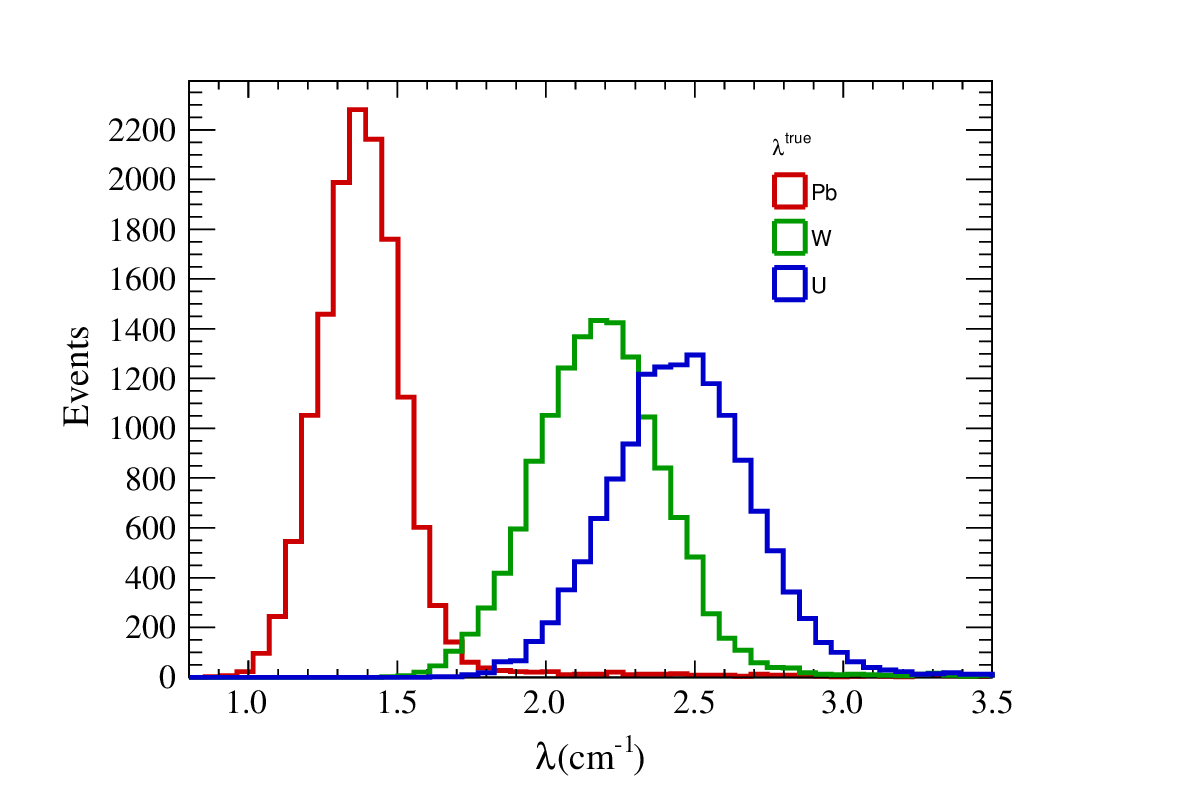}
\end{tabular}
\end{center}
\vspace{-0.5cm}
\caption{The reconstructed scattering density of low-$Z$ and high-$Z$ materials with no momentum measurement in the top row and true momentum information for all muons in the bottom row. 
\label{fig:recLSD36}}
\end{figure}

We start with two cases of no momentum measurement in Eq.~\ref{eq:pocaAveE} and true momentum information for each muon in Eq.~\ref{eq:pocaTrueE} given a spatial resolution of 36 $\mu$m. Their reconstructed scattering density for low-$Z$ and high-$Z$ materials with a large number of independent datasets are shown in Fig.~\ref{fig:recLSD36}. As expected, the high-$Z$ materials and low-$Z$ materials are clearly distinguishable from each other in both scenarios, and precise momentum information improves discrimination capabilities significantly of nuclear material from heavy shielding metals and explosive RDX from low-$Z$ materials. 

\begin{figure}[htbp]
\centering 
\includegraphics[scale=0.5]{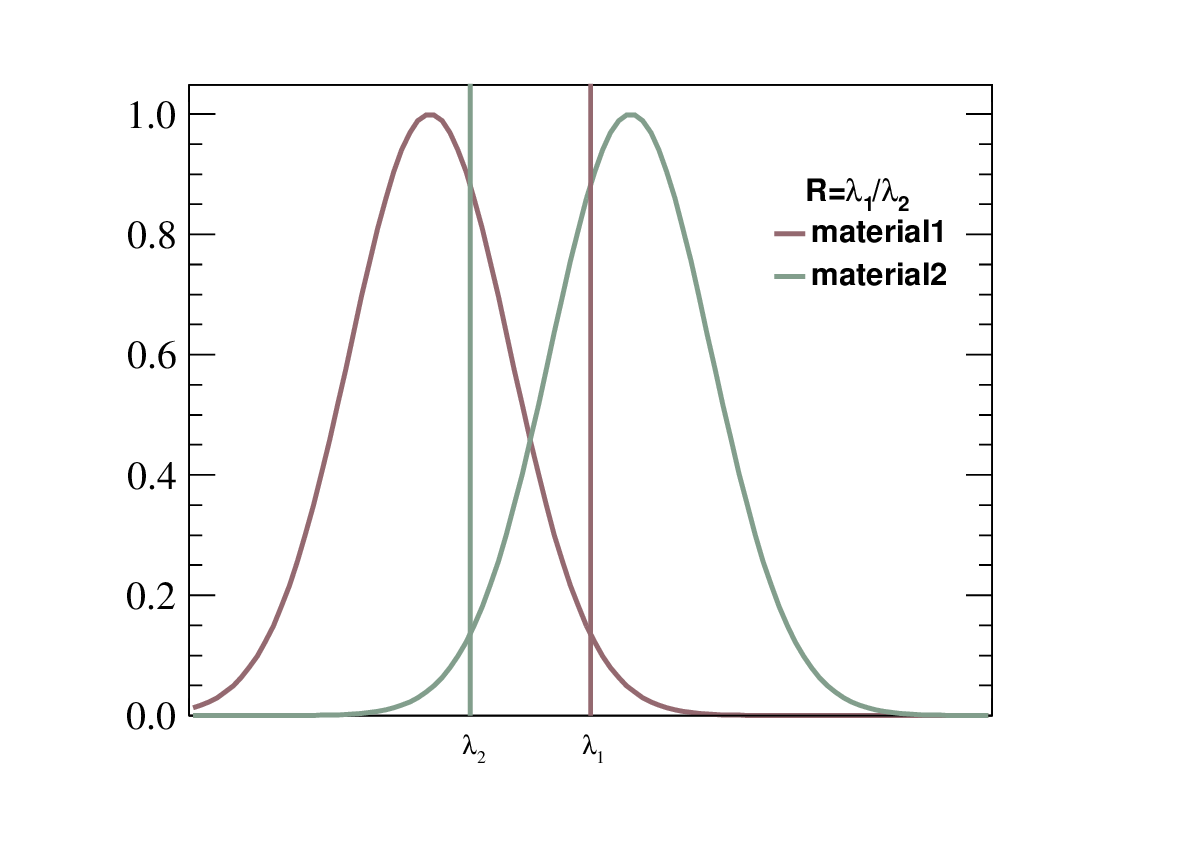}
\vspace{-0.5cm}
\caption{The definition of the ratio $R$ to quantify the separation power of two materials. A smaller value of R represents a better separation power of two materials.
\label{fig:Rdefinition} }
\end{figure}

According to the distributions of reconstructed $\lambda$, we define a ratio $R = \lambda_{1}/\ \lambda_{2}$ to quantify the separation power of two materials~\cite{Benettoni:2013gpa}. As shown in Fig.~\ref{fig:Rdefinition}, $\lambda_{1}$ ($\lambda_{2}$) is the threshold below (above) which the material of a smaller (larger) scattering density could be correctly identified with a probability of $95\%$. Therefore the smaller R-value is, the better separation power of two materials could be achieved. In the following, we evaluate separation powers of explosive RDX in low-$Z$ materials and nuclear material in high-$Z$ materials in the proposed MST system.
\begin{itemize}
\item \textit{\textbf{Effect of LE and HE muons}}: Given a position resolution of 36 $\mu$m and an angular resolution of $3$ mrad, we simulate LE and HE muons in the proposed MST system and reconstruct linear scattering densities with momentum-dependent PoCA method following Eq.~\ref{eq:LSDsegMom}. The separation powers of materials are shown in Fig.~\ref{fig:methodR} with different muon samples. Compared with a typical case of muons without momentum measurement, the LE muons can improve the separation powers of both RDX among low-$Z$ materials and nuclear material among high-$Z$ metals, while HE muons only contribute to the high-Z materials. This is mainly due to very small scattering angles of HE muons in low-$Z$ materials and larger scattering angles of LE muons in a specific material than those of HE muons. 
\begin{figure}[htbp]
\centering 
\includegraphics[scale=0.5]{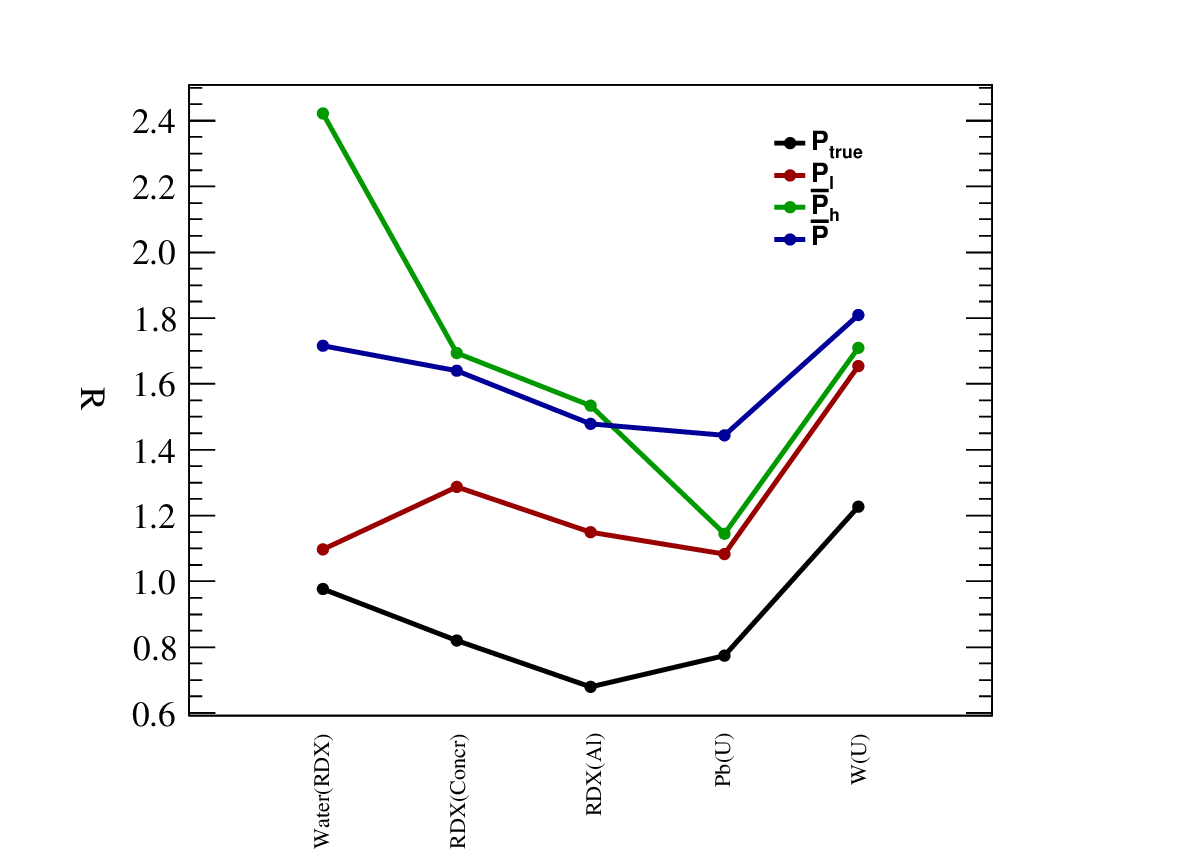}
\caption{The separation powers of materials reconstructed with different muon samples, i.e. muons without momentum measurement, muons with true momentum information, LE muons, and HE muons.
\label{fig:methodR} }
\end{figure}
\item \textit{\textbf{Effect of position resolutions of track detectors}}: In the case of muons with true momentum, the separation powers of materials are obtained with three possible position resolutions of SciFi detector of 36 $\mu$m, 72 $\mu$m and 300 $\mu$m. As shown in the left plot of Fig.~\ref{fig:resR}, a precise track reconstruction of $O(\rm{10 \mu m})$ can improve the low-$Z$ material discrimination significantly than a typical $O(\rm {100 \mu m})$ that is already achieved in micro-pattern gas detectors, but not obviously for the high-$Z$ materials. It is due to the differences of muon deviation angles for two materials as shown in Fig.~\ref{fig:LSDmaterial}, which are about $O(1 \rm{mrad})$ for high-$Z$ materials and $O(0.1 \rm{mrad})$ for low-$Z$ materials in a 5 cm voxel. Therefore $O(\rm{10 \mu m})$ position resolution of track detectors is preferred by low-$Z$ material discrimination and can improve imaging precision of high-$Z$ materials in a smaller voxel size.

\item \textit{\textbf{Effect of angular resolutions of the Cherenkov detector}}: Given a position resolution of 36 $\mu$m in the track detectors, we also check the effect of angular resolutions of the DIRC detector with LE muons. As shown in the right plot of Fig.~\ref{fig:resR}, a realistic resolution of 1.4 mrad and 3 mrad can improve the separation powers for both low-$Z$ and high-$Z$ materials than the average-momentum method. But when an angular resolution gets worse to 5 mrad, the benefit from LE muon momentum may vanish for high-$Z$ materials like W and U due to a smaller fraction and a worse momentum resolution of LE muons. Therefore an angular resolution better than 5 mrad is needed for the DIRC detector.

\begin{figure}[t!]
\vspace{-0.0cm}
\begin{center}
\begin{tabular}{c}
\hspace{-1cm}
\includegraphics[width=0.6\textwidth]{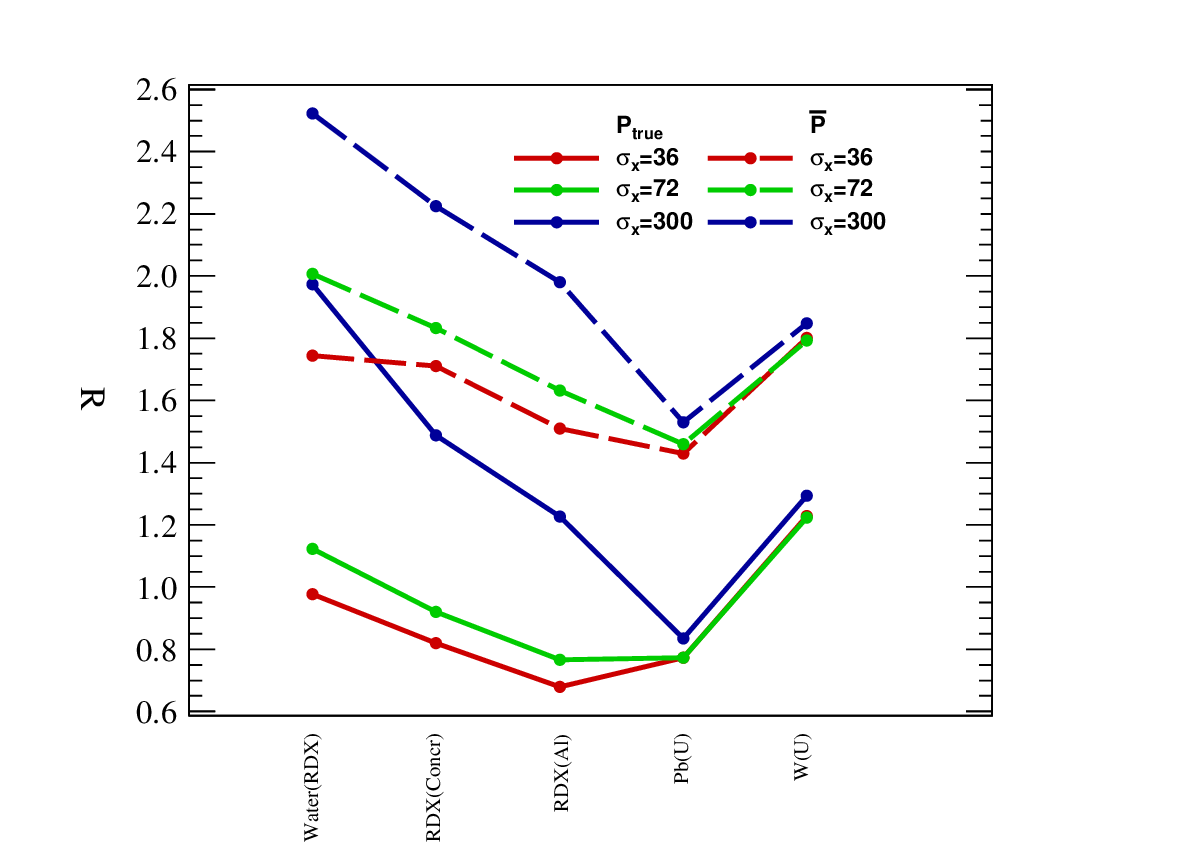}
\hspace{-1.5cm}
\includegraphics[width=0.6\textwidth]{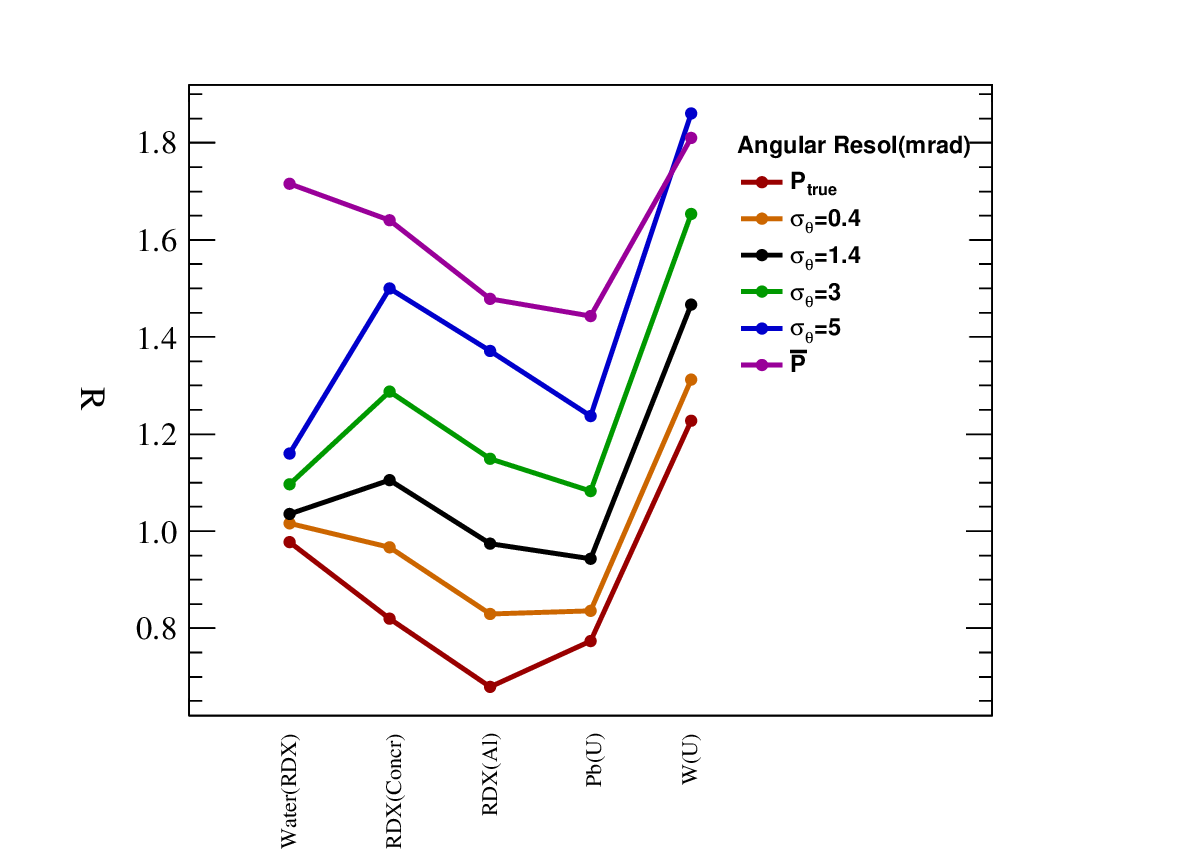}
\end{tabular}
\end{center}
\vspace{-0.5cm}
\caption{The separation powers of different materials in cases of (left) different position resolutions of the SciFi detector with muons of true momentum information or the average momentum and (right) different angular resolutions of the DIRC detector with LE muons given a position resolution of SciFi detector as 36 $\mu$m.
\label{fig:resR}}
\end{figure}
\end{itemize}

\begin{figure}[htbp]
\centering 
\includegraphics[scale=0.5]{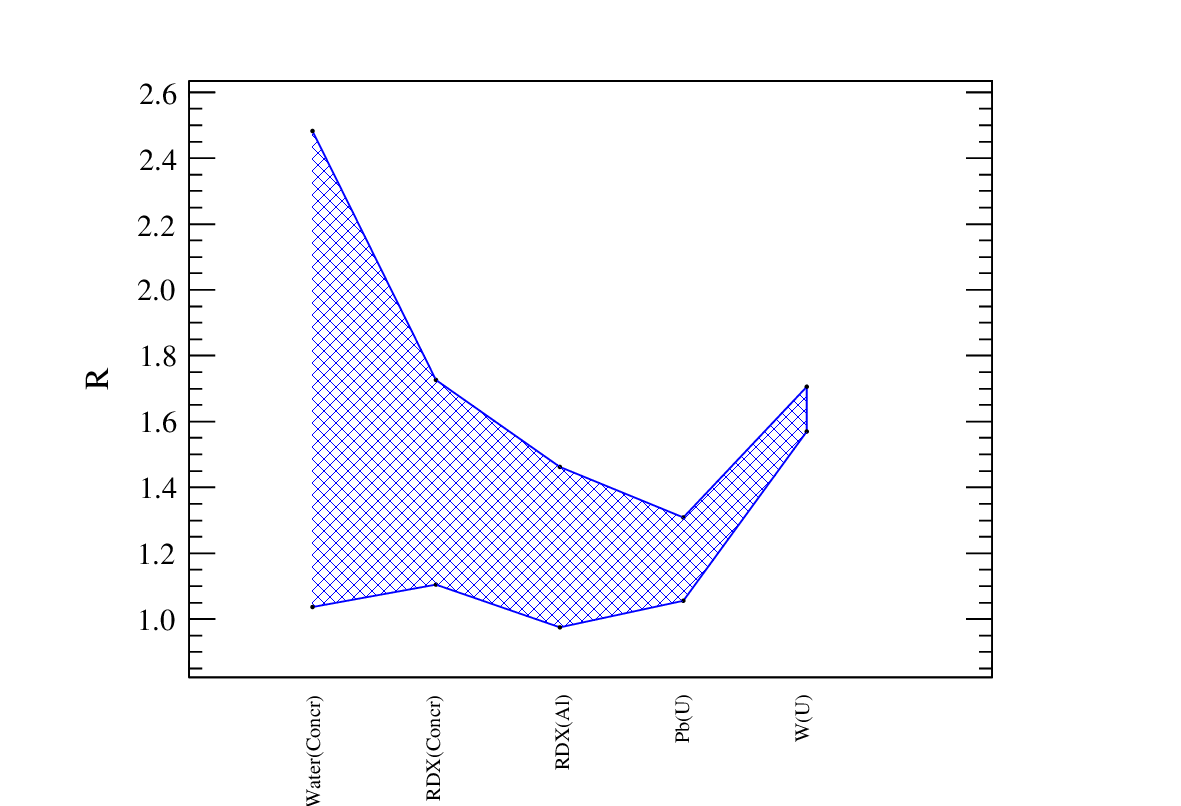}
\caption{The separation power with the realistic position and angular resolution of MST system in the new reconstruction strategy with realistic detector resolutions.}
\label{fig:NewStragy} 
\end{figure}

Based on the results, we suggest a reconstruction strategy in the proposed MST system as follows: firstly the reconstructed voxels are separated into low-$Z$ and high-$Z$ candidates (voxels exclude low-$Z$ candidates) by $\lambda^{ave}$ with all muons. Then the low-$Z$ voxels are reconstructed as $\lambda^{l}$ with LE muons if voxels along their pathlength are all low-$Z$ candidates and high-$Z$ voxels are reconstructed as $\left(\lambda^{l}+\lambda^{h}\right) /\ 2$ with both LE and HE muons for the sake of increasing statistics if at least one of the voxels along their pathlength is a high-$Z$ candidate. The final imaging results can be obtained and optimized by iterating the former step. This strategy is not limited to the PoCA method and also applicable to other advanced algorithms. The separation power of low-$Z$ and high-$Z$ materials with a realistic position resolution of 300 $\mu$m to 36 $\mu$m and a Cherenkov angular resolution of 3 mrad to 1.4 mrad in the new reconstruction strategy is demonstrated in Fig.~\ref{fig:NewStragy} with PoCA algorithm.

\section{Summary and discussion}
In this work, we proposed a solid and compact MST system with extra momentum measurement of low-energy muons by a Cherenkov detector. The MST system is made of two-fold trackers of SciFi detectors and a DIRC detector of fused silica radiator. The SciFi detector has the potential to provide $O(10 {\rm \mu m})$ spatial resolution competitive with traditional silicon detector. The DIRC detector limited by its angular resolution can measure precise momentum of each low-energy muon, but not for high-energy muons, which thus can segment events into LE and HE muons.  We developed a toy Monte Carlo simulation of muon passage through the MST system for statistics tests, which includes a muon generator with a modified Gaisser formula, description of MCS effects in thin sliced material and the detector effects due to spatial resolutions of trackers and angular resolution of DIRC. Since the momentum measured for low-energy muons, a KF algorithm is applied by considering the MCS effect of muons and a segmented momentum-dependent PoCA  method is presented. The full-chain simulated data are implemented for evaluation of material discrimination capabilities of low-$Z$ and high-$Z$ materials in the proposed MST system. Therein the precise reconstructed tracks and the LE muons with precise momentum information can significantly improve the separation power of explosive RDX among low-$Z$ materials and nuclear material among high-$Z$ materials compared with the typical average-momentum method. In the end, we proposed a reconstruction strategy with LE and HE muons for both low-$Z$ and high-$Z$ muons in the proposed MST system.

The key point to achieve a good separation power of low-$Z$ and high-$Z$ materials in this proposed MST system is a high position resolution of track detectors and accurate momentum measurement of low energy muons. Since not enough information are provided in other state-of-art MST systems~\cite{CRTBorderSec:2023} to directly compare with this proposed system, we make a discussion qualitatively here according to the detector structures and resolutions:
\begin{itemize}
\item Currently a position resolution of the track detector better than 100$\rm{\mu m}$ has already been achievable in a MST system with gas detectors, such as 85 $\rm{\mu m}$ of Micromegas~\cite{Wang:2021ssd} and 50 $\rm{\mu m}$ of GEM detetor~\cite{Gnanvo:2011}. The main advantage of the SciFi tracker lies in its solid and compact structures, which can be portable and more robust in real applications.

\item Regarding the muon momentum measurement, there are mainly two feasible strategies by inserting additional absorption layers of known materials~\cite{Anghel:2015} or adding a fieldable Cherenkov detector~\cite{Bae:2022dti,Bae:2022} for a MST system besides this work. For the former one, a good momentum measurement of low energy muons needs multiple plates inserted~\cite{Oláh:2022}, which results in a quite large-size muon spectrometer than the DIRC detector. For the latter one, Ref.~\cite{Bae:2022} shows that the fieldable Cherenkov detector can measure muon momentum with a resolution of $\pm0.5 {\rm GeV/c}$ within a momentum range of 0.1 to 10.0 $\rm{GeV/c}$, which is worse than the DIRC detector for muons with energy below about 1.5 GeV given a $\sigma_{\theta}$ as 1.4 mrad, but better for muons with higher energies. It means that the DIRC detector can get a better discrimination of low-$Z$ materials. Additionally, the DIRC detector is solid and compact, which can be portable and more robust in real applications together with SciFi detectors. 
\end{itemize}

We ignored the energy loss of cosmic ray muons crossing through materials in this work. For small and medium sized objects like passenger luggage, the energy loss is not serious and the measured LE and HE muons in the Cherenkov detectors are still feasible in the suggested reconstruction strategy for identification of both low-$Z$ and high-$Z$ materials. However, for large sized objects, part of low-energy muons might be absorbed, and the momentum measured by the DIRC detector of survival muons can be lower than that before the target volume. There are two possible ways to compensate reconstruction performance that one is to include absorbed muons selected by upper and lower trackers in the reconstruction like Ref.~\cite{Blanpied:2015,Anbarjafari:2021} and the other is to estimate the energy loss along the path length of a survival muon and add it into the reconstruction process as constrains. All these shall be studied in the latter studies with detailed simulations of MST system, such as in Geant4 software.

\section{Acknowledgment}
This work was supported in part by the National Natural Science Foundation of China under Grant No. 12205174, and by Shandong Provincial Natural Science Foundation under Grant No. ZR2021QA097 and No. ZR2021QA107.

\end{document}